%% file: EHI_band.tex
\documentclass[11pt,letterpaper]{article}
\pdfoutput=1
\usepackage{jcappub}
\usepackage{ulem}
\usepackage{array}
\title{Extended Maximum Likelihood Halo-independent Analysis of Dark Matter Direct Detection Data}

\author[a]{Graciela B. Gelmini,}
\author[a]{Andreea Georgescu,}
\author[b]{Paolo Gondolo,}
\author[a]{and Ji-Haeng Huh}

\affiliation[a]{Department of Physics and Astronomy, UCLA,\\
475 Portola Plaza, Los Angeles, CA 90095, USA}
\affiliation[b]{Department of Physics and Astronomy, University of Utah,\\
115 South 1400 East \#201, Salt Lake City, UT 84112, USA}

\emailAdd{gelmini@physics.ucla.edu}
\emailAdd{a.georgescu@physics.ucla.edu}
\emailAdd{paolo@physics.utah.edu}
\emailAdd{jhhuh@physics.ucla.edu}

\newcommand{\vmin}{v_\text{min}}
\newcommand{\bsv}{\boldsymbol{v}}

\abstract{
We extend and correct a recently proposed maximum-likelihood halo-independent method to analyze unbinned direct dark matter detection data.
Instead of the recoil energy as independent variable we use the minimum speed a dark matter particle must have to impart a given recoil energy to a nucleus.
This has the advantage of allowing us to apply the method to any type of target composition and interaction, e.g.~with general momentum and velocity dependence, and with elastic or inelastic scattering.
We prove the method and provide a rigorous statistical interpretation of the results.
As first applications,
we find that for dark matter particles with elastic spin-independent interactions and neutron to proton coupling ratio $f_n/f_p=-0.7$, the WIMP interpretation of the signal observed by CDMS-II-Si is compatible with the constraints imposed by all other experiments with null results.
We also find a similar compatibility for exothermic inelastic spin-independent interactions with $f_n/f_p=-0.8$.
}


\begin{document}

\maketitle

\setcounter{page}{1}
 

\section{Introduction} \label{sec:introduction}

We do not know what the dark matter (DM), the most abundant form of matter in the universe, consists of.
Weakly interacting massive particles (WIMPs) are the most extensively studied DM particle candidates, not only because of their theoretical appeal but also because they could be detected in the near future.

Direct searches attempt to measure the energy WIMPs might deposit when interacting within a detector.
Three direct search experiments, DAMA \cite{Bernabei:2010mq}, CoGeNT \cite{Aalseth:2010vx,Aalseth:2012if,Aalseth:2011wp,Aalseth:2014eft,Aalseth:2014jpa}, and CDMS-II-Si \cite{Agnese:2013rvf} have potential DM signals, while all the other searches have produced only upper bounds on scattering rates and their annual modulation \cite{Angle:2011th,Aprile:2011hi,Aprile:2012nq,Ahmed:2010wy,Felizardo:2011uw,Archambault:2012pm,Behnke:2012ys,Ahmed:2012vq,Agnese:2013lua,Akerib:2013tjd,Agnese:2014aze}.

The conventional analysis of direct search data relies on a specific model of the DM halo of our galaxy, often chosen to be the standard halo model (SHM).
A halo-independent data comparison method has been developed more recently \cite{Fox:2010bz,Frandsen:2011gi,Gondolo:2012rs,Frandsen:2013cna,DelNobile:2013cta,DelNobile:2013cva}.
The basic idea behind this method is that all the dependence of the scattering rate on the halo model, in any detector, resides in the same function which we call $\tilde\eta(\vmin,t)$ of the speed $\vmin$ and the time $t$.

Since $\tilde\eta(\vmin,t)$ is common to all direct search experiments, this function can be measured by all experiments, and the compatibility of the different measurements can be studied.
The speed $v_{\rm min}$ is the minimum speed necessary for the incoming interacting DM particle to impart a recoil energy $E_R$ to a nucleus in each detector. 
Conversely, given an incoming WIMP speed $v=\vmin$, $E_R$ is the extremum recoil energy (maximum energy for elastic collisions, or either maximum or minimum for inelastic collisions) that the DM particle can impart to a nucleus.
Notice that $E_R$ and $v_{\rm min}$ are exchangeable variables only for a single nuclide. When a target consists of multiple nuclides, a choice must be made between the two, $E_R$ and $v_{\rm min}$.
Taking $E_R$ as independent variable (as is done in \cite{Fox:2010bz,Frandsen:2011gi,Frandsen:2013cna}) $v_{\rm min}$ depends on each target nuclide.
In our approach, $v_{\rm min}$ and the observed energy $E'$ are the independent variables.
This allows us to incorporate any isotopic composition of the target by summing over target nuclide dependent $E_R(v_{\rm min})$ for fixed observed $E'$.

In earlier implementations of the halo-independent method, only weighted averages over $\vmin$ intervals of the time average, $\tilde\eta^0(\vmin)$, and annual modulation amplitudes, $\tilde\eta^1(\vmin)$, of $\tilde\eta(\vmin,t)$ have been obtained from putative DM signals in direct detection.
These averages over $\vmin$ intervals are represented in plots by a set of crosses in the $\vmin-\tilde\eta$ plane, whose vertical and horizontal bars show the uncertainty in $\tilde\eta^0(\vmin)$ or $\tilde\eta^1(\vmin)$ and the $\vmin$ range where they are measured, respectively.
Combined with upper limits, these crosses can be used to assess the compatibility of data sets from various experiments.
However, making a statistically meaningful evaluation of the compatibility of the data in this manner is not possible.

The compatibility of different data has been studied in \cite{Feldstein:2014ufa} using the ``parameter goodness-of-fit'' test statistic \cite{Maltoni:2003cu}. The analysis is based on the likelihoods maximized with $\tilde\eta^0$ written as a sum of very large number of step functions, following a method presented in \cite{Feldstein:2014gza}.
In this case, the level of compatibility is given by the p-value of the test statistic, which was calculated by Monte Carlo simulations in \cite{Feldstein:2014ufa}.
Another test statistic for comparing one data set with a positive result and another with a negative result has been defined in \cite{Bozorgnia:2014gsa}.

An alternative method to study the compatibility of a positive result with upper limits uses a band in $v_{\rm min}-\tilde\eta^0$ space at a given confidence level \cite{Fox:2014kua}, derived from unbinned data, with an extended likelihood \cite{Barlow:1990vc}. 
In this case, as shown in \cite{Fox:2014kua} for single-nuclide detectors, the likelihood is maximized by a non-increasing piecewise constant $\tilde\eta^0$ function, because of the exponential prefactor in the extended likelihood.

The proof presented in \cite{Fox:2014kua} relies on the assumption that the target is made of a single component.
The main limitation of the approach of \cite{Fox:2014kua} relies on their use of the recoil energy $E_R$ as independent variable.
Here we provide a derivation of the extended maximum likelihood halo-independent (EHI) analysis method using $v_{\rm min}$ as a variable, which applies to any type of WIMP interaction, including inelastic scattering, and any target composition.
We correct and extend the original proof of \cite{Fox:2014kua} using the formulation developed for the generalized halo-independent analysis in \cite{DelNobile:2013cva}.
The proof for the realistic case of finite experimental energy resolution presented in \cite{Fox:2014kua} relies on the application to the likelihood functional maximization of the Karush-Kuhn-Tucker (KKT) conditions in \cite{Karush,KuhnTucker}.
The KKT conditions in \cite{Karush,KuhnTucker}, however, apply only to the minimization of functions with a finite number of variables subject to a finite number of inequalities, and they do not apply to functionals.
Eqs.~(A.3) to (A.6) of \cite{Fox:2014kua} are given without proof and without a reference. 
Moreover, Eq.~(A.4) seems problematic for a $\tilde g$ function (which in our paper we call $\tilde\eta$) that has discontinuities, as in the solutions found in \cite{Fox:2014kua}. In this case, Eq.~(A.4) requires a Dirac $\delta$ function to be smaller or equal to zero, which is mathematically problematic.
As we explain in Sec.~3, although the KKT conditions have been extended to functionals defined on specific kinds of function spaces and constraints, we did not find in the literature a proof that clearly applies to our problem.
Thus, in Sec.~3 we present our own proof of the KKT conditions we use, Eqs.~\eqref{eq:I}-\eqref{eq:IV}, which are clearly valid for discontinuous functions. 

As in \cite{Fox:2014kua}, here we find that the best fit $\tilde\eta$ function is piecewise constant with a number of discontinuities at most equal to the number of observed events. 
In \cite{Fox:2014kua}, this is a result found for $\tilde g$ given as a function of the recoil energy, which can be easily translated to $\vmin$ space only for a single target nuclide. 
Besides, the proof in \cite{Fox:2014kua} applies only to resolution functions with certain properties.
We instead prove the result for $\tilde\eta$ as a function of $\vmin$ for any target composition and general resolution functions.

Besides these extensions, we make a correction to the method of \cite{Fox:2014kua} by providing a clear definition of the uncertainty band.
In \cite{Fox:2014kua}, the uncertainty band is defined in Eq.~(2.16) through a numerical Monte Carlo simulation. 
In Sec.~4, we explain our objections to this procedure. 
We define instead a pointwise confidence band with a new method (see Sec.~4.2) and provide a clear statistical interpretation for this band using Wilks' theorem. 
The different definitions of the band, here and in \cite{Fox:2014kua}, yield very different values of the parameter $\Delta L$ defined in both papers for the same confidence level.

In Sec.~\ref{sec:genHI}, we review the formulation of the generalized halo-independent analysis, on which the following sections are based.
In Sec.~\ref{sec:ubHI}, we prove crucial properties of the extended likelihood for unbinned direct dark matter detection data.
In Sec.~\ref{sec:EHI}, we develop the EHI analysis method, and discuss the statistical interpretation of the confidence band computed with this method.
In Sec.~\ref{sec:apps}, we apply the method to the CDMS-II-Si \cite{Agnese:2013rvf} data for WIMPs with elastic isospin-conserving and isospin-violating SI interactions \cite{Kurylov:2003ra, Chang:2010yk, Feng:2011vu}, and exothermic inelastic isospin-violating SI interactions \cite{Gelmini:2014psa,Scopel:2014kba}, and compare the results with the upper limits imposed by other experiments.
Finally, we give our conclusions in Sec.~\ref{sec:con}.

\section{Generalized halo-independent analysis method}\label{sec:genHI}

The differential recoil rate per unit detector mass, typically given in units of $\text{counts}/\text{day}/\text{kg}/\text{keV}$, for the scattering of WIMPs of mass $m$ off a target nuclide $T$ with mass $m_T$ is
\begin{equation} 
\frac{{\rm d}R_T}{{\rm d}E_R} 
=\frac{\rho}{m} \frac{C_T}{m_T}
\int_{v\geq \vmin(E_R)}{\rm d}^3v~f(\bsv,t)v
\frac{{\rm d}\sigma_T}{{\rm d}E_R} (E_R,\bsv),
\label{eq:dRdE}
\end{equation}
where $C_T$ is the mass fraction of nuclide $T$ in the detector, $E_R$ is the nuclear recoil energy, $\rho$ is the WIMP local energy density, $f(\bsv,t)$ is the WIMP velocity distribution in Earth's frame, ${\rm d}\sigma_T/{\rm d}E_R$ is the WIMP-nucleus differential scattering cross section, and $v_\text{min}(E_R)$ is the minimum WIMP speed needed to impart to the target nucleus a recoil energy $E_R$.
The revolution of Earth around the Sun introduces an annual modulation of $f(\bsv,t)$.
In detectors with more than one nuclide in their target, the total differential recoil rate is
\begin{equation} 
\frac{{\rm d}R}{{\rm d}E_R} 
=\sum_T\frac{{\rm d}R_T}{{\rm d}E_R}.
\end{equation}

Allowing for the possibility of inelastic scattering of an incoming WIMP with mass $m$ into another outgoing WIMP with mass $m'=m+\delta$, for $\mu_T|\delta|/m^2 \ll 1$, $v_{\rm min}(E_R)$ is given by
\begin{equation}
\vmin(E_R)
=\frac{1}{\sqrt{2m_TE_R}}
\left|\frac{m_TE_R}{\mu_T}+\delta\right|,
\label{eq:vminER}
\end{equation}
where $\mu_T\equiv (m\:m_T)/(m+m_T)$ is the WIMP-nucleus reduced mass.
By inverting this equation one obtains the minimum and maximum recoil energies $E_R^{T,\pm}(v)$ that are kinematically allowed for a fixed DM speed $v$,
\begin{equation}
E_R^{T,\pm}(v) =
\frac{\mu_T^2 v^2}{2 m_T} \left( 1 \pm \sqrt{1 - \frac{2 \delta}{\mu_T v^2}} \right)^2 \ .
\label{eq:ER_vs_v}
\end{equation}
The minimum possible value of $\vmin$, thus also of $v$, for the interaction to be kinematically allowed is $v^T_\delta=\sqrt{2\delta/\mu_T}$ for endothermic scattering, and $v^T_\delta = 0$ for elastic and exothermic scattering.
This speed value corresponds to the point of intersection of the two $E_R^{T,\pm}$ branches.
We define $v_\delta$ to be the smallest of the $v_\delta^T$ values among all nuclides $T$ in the detector.

Most experiments do not measure the recoil energy $E_R$ directly.
They measure instead a proxy $E'$ for it, such as the ionization or scintillation signals, subject to experimental uncertainties and fluctuations.
These are represented by an energy response function $G_T(E_R,E')$, which is the probability distribution for an event with recoil energy $E_R$ to be measured with energy $E'$.
Including the experimental acceptance $\epsilon(E_R,E')$ from various experimental cuts and efficiencies, which, in general, is a function of both the recoil energy $E_R$ and the detected energy $E'$, the differential event rate in the detected energy $E'$ can be written as
\begin{equation}
\frac{{\rm d}R}{{\rm d}E'}
=\sum_T \int_0^\infty{\rm d}E_R
~\epsilon(E_R,E') G_T(E_R,E')\frac{{\rm d}R_T}{{\rm d}E_R}.
\label{eq:dRdEE}
\end{equation}

By inserting
\eqref{eq:dRdE} into \eqref{eq:dRdEE} we can express the differential event rate in the detected energy $E'$ as a double integral
\begin{eqnarray}
\frac{{\rm d}R}{{\rm d}E'}
&=&\frac{\rho}{m}\sum_T
\frac{C_T}{m_T}\int_0^\infty{\rm d}E_R
~\epsilon(E_R,E') G_T(E_R,E')\nonumber\\
&&\times\int_{v\geq \vmin(E_R)}{\rm d}^3v
~f(\bsv,t)v\frac{{\rm d}\sigma_T}{{\rm d}E_R} (E_R,\bsv),
\end{eqnarray}
from where changing the integration order and extracting a reference parameter $\sigma_{\rm ref}$ from the cross section we get
\begin{eqnarray}
\frac{{\rm d}R}{{\rm d}E'}
&=&\frac{\sigma_{\rm ref}\rho}{m}
\int_{v\ge v_\delta}{\rm d}^3v
~\frac{f(\bsv,t)}{v}\frac{{\rm d}{\cal H}}{{\rm d}E'}(E',\bsv).
\label{eq:dRdE'}
\end{eqnarray}
Here the function ${\rm d}{\cal H}/{\rm d}E'$ is
\begin{equation}
\frac{{\rm d}{\cal H}}{{\rm d}E'}(E',\bsv)
\equiv \sum_T
\frac{{\rm d}{\cal H}_T}{{\rm d}E'}(E',\bsv),
\label{eq:calH}
\end{equation}
where ${\rm d}{\cal H}_T/{\rm d}E'$ is defined as
\begin{equation}
\frac{{\rm d}{\cal H}_T}{{\rm d}E'}(E',\bsv)
\equiv 
\left\{
  \begin{array}{>{\displaystyle}l>{\displaystyle}l}
&\frac{C_T}{m_T}\int_{E_R^{T,-}(v)}^{E_R^{T,+}(v)}{\rm d}E_R
~\epsilon(E_R,E') G_T(E_R,E')\frac{v^2}{\sigma_{\rm ref}}
\frac{{\rm d}\sigma_T}{{\rm d}E_R} (E_R,\bsv) ~~~\text{ if } v\ge v_\delta^T,\\
&0  ~~~ \text{ if } v<v_\delta^T.
  \end{array}
\right.
\label{eq:calHT}
\end{equation}

In \eqref{eq:dRdE'} and \eqref{eq:calHT} we have written explicitly a parameter $\sigma_{\rm ref}$ extracted from the differential cross section to represent the strength of the interaction.
This is preferably, but not necessarily, the WIMP-proton scattering cross section.

Here we consider only differential cross sections (and thus also ${\rm d}{\cal H}/{\rm d}E'$ functions) which depend only on the speed $v=|\bsv|$, and not on the direction of the initial WIMP velocity $\bsv$.
The cross section depends only on $v$ if the incoming WIMPs and the target nuclei are unpolarized and the detector response is isotropic, as is most common.
In this case, one can write the differential event rate in a simpler form as
\begin{equation}
\frac{{\rm d}R}{{\rm d}E'}
=\frac{\sigma_{\rm ref}\rho}{m}
\int_{v_\delta}^\infty{\rm d}v~\frac{F(v,t)}{v}
\frac{{\rm d}{\cal H}}{{\rm d}E'}(E',v),
\label{eq:dRdE'iso}
\end{equation}
where $F(v,t)\equiv v^2\int{\rm d}\Omega_v~f(\bsv,t)$.
We now define the function $\tilde\eta(\vmin,t)$ as
\begin{equation}
\tilde\eta(\vmin,t)
\equiv\frac{\rho\sigma_{\rm ref}}{m}
\int_{\vmin}^\infty{\rm d}v~\frac{F(v,t)}{v}.
\label{eq:eta}
\end{equation}
Thus,
\begin{equation}
\frac{\sigma_{\rm ref}\rho}{m}\frac{F(v,t)}{v}
=-\frac{\partial\tilde\eta(v,t)}{\partial v},
\end{equation}
and \eqref{eq:dRdE'iso} becomes
\begin{eqnarray}
\frac{{\rm d}R}{{\rm d}E'}
&=&-\int_{v_\delta}^\infty{\rm d}v~
\frac{\partial\tilde\eta(v,t)}{\partial v}
\frac{{\rm d}{\cal H}}{{\rm d}E'}(E',v).
\label{eq:dRdE'partialEta}
\end{eqnarray}
Using that $\tilde\eta(\infty,t)=0$ (see \eqref{eq:eta}) and ${\rm d}{\cal H}/{\rm d}E'(E',v_\delta)=0$ (since $E_R^{T,-}(v_\delta)= E_R^{T,+}(v_\delta)$ and the integrand in \eqref{eq:calH} is a regular function), the integration by parts of \eqref{eq:dRdE'partialEta} leads to
\begin{eqnarray}
\frac{{\rm d}R}{{\rm d}E'}
&=&\int_{v_\delta}^\infty{\rm d}\vmin~
\tilde\eta(\vmin,t)
\frac{{\rm d}{\cal R}}{{\rm d}E'}(E',\vmin),
\label{eq:rateInCalR}
\end{eqnarray}
where we choose to call $\vmin$ the integration variable because it makes obvious the physical meaning of $\tilde\eta$ as a function of $\vmin$, and where we define the ``differential response function'' ${\rm d}{\cal R}/{\rm d}E'$ of the detector as
\begin{equation}
\frac{{\rm d}{\cal R}}{{\rm d}E'}(E',\vmin)\equiv\frac{\partial}{\partial \vmin}\left[\frac{{\rm d}{\cal H}}{{\rm d}E'}(E',\vmin)\right].
\label{eq:calR}
\end{equation}
Notice that ${\rm d}{\cal R}/{\rm d}E'$ is a function of the target dependent recoil energies, $ E_R^{T,\pm}(\vmin)$, which are functions of the independent variable $\vmin$.
It is clear that, in \eqref{eq:rateInCalR}, all the dependence on the halo model is in the $\tilde\eta$ function which is independent of the experimental apparatus, and thus is common to all direct detection experiments.
Therefore, by mapping the rate data into $\tilde\eta$, it is possible to compare the different experimental results without any assumption on the dark halo of our galaxy.

With ${\rm d}R/{\rm d}E'$ given in \eqref{eq:rateInCalR}, the energy-integrated rate over an energy interval $[E'_1,E'_2]$ is
\begin{equation}
R_{[E'_1,E'_2]}(t)\equiv\int_{E'_1}^{E'_2}{\rm d}E'~\frac{{\rm d}R}{{\rm d}E'}.
\end{equation}
Direct detection experiments can measure the time-average $R^0_{[E'_1,E'_2]}$ and the annual modulation amplitude $R^1_{[E'_1,E'_2]}$ of the rate, $R_{[E'_1,E'_2]}\simeq R^0_{[E'_1,E'_2]}+R^1_{[E'_1,E'_2]}\cos(2\pi(t-t_0)/{\rm yr})$, with phase $t_0$.
The only source of time dependence in the rate is $\tilde\eta$, thus
\begin{eqnarray} 
R^\alpha_{[E'_1,E'_2]}
&\equiv& \int_{v_\delta}^{\infty}{\rm d}\vmin~
\tilde\eta^\alpha(\vmin)
\int_{E'_1}^{E'_2}{\rm d}E'~
\frac{{\rm d}{\cal R}}{{\rm d}E'}\nonumber\\
&=&\int_{v_\delta}^{\infty}{\rm d}\vmin~\tilde\eta^\alpha(\vmin)
{\cal R}_{[E'_1,E'_2]}(\vmin)
\end{eqnarray} 
where $\alpha=0$ or $1$, and $\tilde\eta(v,t)\simeq\tilde\eta^0(v)+\tilde\eta^1(v)\cos(2\pi(t-t_0)/{\rm yr})$.

If the energy-integrated response function ${\cal R}_{[E'_1,E'_2]}(\vmin)$ for a given energy interval $[E'_1,E'_2]$,
\begin{equation}\label{eq:energy_integr_resp}
  {\cal R}_{[E'_1,E'_2]}(\vmin)=\int_{E'_1}^{E'_2}{\rm d}E'\frac{{\rm d}{\cal R}}{{\rm d}E'}(E',\vmin),
\end{equation}
is a well-localized function in a $\vmin$ range, the measurements of $R^0_{[E'_1,E'_2]}$ and $R^1_{[E'_1,E'_2]}$ can be used to infer the values of $\tilde\eta^0$ and $\tilde\eta^1$ over the $\vmin$ range in which the response function is non-zero.
This is true for WIMPs whose differential cross section is inversely proportional to $v^2$, such as for the usual spin-independent (SI) and spin-dependent interactions.
Otherwise ${\cal R}_{[E'_1,E'_2]}$ may need to be regularized (see \cite{DelNobile:2013cva} for the details).
 
For SI interactions, the WIMP-nucleus differential cross section can be written in terms of the effective couplings of the WIMP with neutrons and protons, $f_n$ and $f_p$, as
\begin{equation} 
\frac{{\rm d}\sigma^{SI}_T}{{\rm d}E_R}(E_R,v)
=\sigma_p \frac{\mu_T^2}{\mu_p^2} [Z_T+ (A_T-Z_T)(f_n/f_p)]^2~\frac{F_T^2(E_R)}{2\mu_T^2v^2/m_T},
\label{eq:sigmaSI}
\end{equation}
where $\sigma_p$ is the WIMP-proton cross section, $\mu_p$ is the WIMP-proton reduced mass, $A_T$ and $Z_T$ are the atomic and charge numbers of the nuclide $T$, respectively, and $F_T^2(E_R)$ is a nuclear form factor, for which we take the Helm form factor~\cite{Helm:1956zz} normalized to $F_T^2(0) = 1$.
Using \eqref{eq:sigmaSI} in \eqref{eq:calH} we obtain
\begin{eqnarray}
\frac{{\rm d}{\cal H}^{\rm SI}}{{\rm d}E'} (v,E')
&\equiv&
\sum_T
\frac{C_T}{2\mu_p^2}
\int_{E_R^{T,-}(v)}^{E_R^{T,+}(v)}{\rm  d}E_R~
\epsilon(E_R,E')G_T(E_R,E')\nonumber\\
&&\times
   [Z_T+(A_T-Z_T)(f_n/f_p)]^2F_T^2(E_R),
\end{eqnarray}
and from \eqref{eq:calR}, we get the following differential response function:
${\rm d}{\cal R}^{\rm SI}/{\rm d}E'$
\begin{eqnarray}
\frac{{\rm d}{\cal R}^{\rm SI}}{{\rm d}E'} (\vmin,E')
&\equiv&
\sum_T
\frac{C_T}{2\mu_p^2}
[Z_T+(A_T-Z_T)(f_n/f_p)]^2\nonumber\\
&&\times \left[
\left.\frac{{\rm d}E_R^{T,+}}{{\rm d} v}\right|_{v=\vmin}
\epsilon(E_R^{T,+}(\vmin),E')G_T(E_R^{T,+}(\vmin),E')F_T^2(E_R^{T,+}(\vmin))
\right.\nonumber\\
&&
\left.
-
\left.\frac{{\rm d}E_R^{T,-}}{{\rm d} v}\right|_{v=\vmin}
\epsilon(E_R^{T,-}(\vmin),E')G_T(E_R^{T,-}(\vmin),E')F_T^2(E_R^{T,-}(\vmin)) 
\right]. \nonumber\\
\label{eq:calRSI}
\end{eqnarray}
For elastic scattering this reduces to
\begin{eqnarray}
\frac{{\rm d}{\cal R}^{\rm SI}}{{\rm d}E'} (\vmin,E')
&\equiv&\sum_T2\vmin\frac{C_T}{m_T} \epsilon(E_R,E')G_T(E_R(\vmin),E')\nonumber\\
&&\times\frac{\mu_T^2}{\mu_p^2}
   [Z_T+(A_T-Z_T)(f_n/f_p)]^2F_T^2(E_R(\vmin)).
\end{eqnarray}

\section{Piecewise constant $\tilde\eta(v_{\rm min})$ resulting from the EHI method}\label{sec:ubHI}

Most direct detection experiments measure energy-integrated rates and/or their annual modulation amplitudes in given energy intervals.
CDMS-II-Si gives instead the recoil energies of three candidate DM events.
Most halo-independent analyses of the CDMS-II-Si candidate events have chosen a binning scheme, which is arbitrary and may lose some of the information in the data \cite{Frandsen:2013cna, DelNobile:2013cta, DelNobile:2013cva, DelNobile:2014eta, Gelmini:2014psa, Scopel:2014kba, DelNobile:2014sja}.

Ref.~\cite{Fox:2014kua} has introduced a halo-independent analysis method without binning.
The method relies on the fact that the extended likelihood \cite{Barlow:1990vc} yields piecewise constant functions as solutions of the likelihood maximization.
The extended likelihood for unbinned data can be written as
\begin{equation}
{\cal L}[\tilde\eta(\vmin)]
\equiv e^{-N_E[\tilde\eta]} \prod_{a=1}^{N_O}MT\left.\frac{{\rm d}R_{tot}}{{\rm d}E'}\right|_{E'=E'_a}.
\label{eq:extendedLikelihood}
\end{equation}
For simplicity we use $\tilde\eta$ here for the time-average component of the $\tilde\eta$ function (we call it $\tilde\eta^0$ in previous sections).
Here $N_O$ is the total number of observed events, each with energy $E'_a$, with $a=1,\dots,N_O$.
$N_E[\tilde\eta]$ is the total number of expected events within the energy range $[E'_{\rm min},E'_{\rm max}]$ detectable in the experiment, which we write as a functional of the function $\tilde\eta(\vmin)$:
\begin{equation}\label{eq:NE}
N_E[\tilde\eta]=N_{BG} + MT\int_{v_\delta}^\infty {\rm d}\vmin\tilde\eta(\vmin){\cal R}_{[E'_{\rm min},E'_{\rm max}]}(\vmin),
\end{equation}
where $N_{BG}$ is the expected number of background events
\begin{equation}
N_\text{BG} \equiv MT\int_{E'_\text{min}}^{E'_\text{max}} dE' \dfrac{dR_\text{BG}}{dE'}.
\label{eq:NBG}
\end{equation}
Here $MT$ is the detector exposure, ${\rm d}R_{tot}/{\rm d}E'$ is the total predicted differential event rate
\begin{eqnarray}
\frac{{\rm d}R_{tot}}{{\rm d}E'}&=&\frac{{\rm d}R_{\rm BG}}{{\rm d}E'}+\frac{{\rm d}R}{{\rm d}E'}\nonumber\\
&=&\frac{{\rm d}R_{\rm BG}}{{\rm d}E'}
+\int_{v_\delta}^\infty {\rm d}\vmin\tilde\eta(\vmin)\frac{{\rm d}{\cal R}}{{\rm d}E'}(\vmin),
\label{eq:dRtot_dE}
\end{eqnarray}
and ${\rm d}R_{\rm BG}/{\rm d}E'$ is the differential rate of the background events.
Writing the rate in this form allows to take into account a non-trivial target composition (not included in \cite{Fox:2014kua}), through the differential response function ${\rm d}{\cal R}/{\rm d}E'$, defined in \eqref{eq:calH} and \eqref{eq:calR}, or in \eqref{eq:calRSI} for SI interactions.

Without fixing the halo model, the likelihood function in \eqref{eq:extendedLikelihood} is actually a functional of the $\tilde\eta$ function.
If there is no uncertainty in the measurement of recoil energies, for a single target nuclide it was proven in \cite{Fox:2014kua} that the likelihood is maximized by a piecewise constant $\tilde\eta$ function with the number of steps equal to or smaller than the number of observed events, $N_O$.
The proof for the realistic case with a finite energy resolution presented in \cite{Fox:2014kua} applies only to resolution functions with certain properties such as having a single local maximum, and relies on the application to the likelihood functional maximization of the Karush-Kuhn-Tucker (KKT) conditions in \cite{Karush,KuhnTucker}.
The KKT conditions in \cite{Karush,KuhnTucker} apply to the minimization of functions with a finite number of variables subject to a finite number of inequalities.
The proofs in \cite{Karush,KuhnTucker} do not apply to functionals.
The likelihood ${\cal L[\tilde\eta]}$ in \eqref{eq:extendedLikelihood} is instead a functional of $\tilde\eta(\vmin)$ subject to an infinite number of inequalities, one for each value of $\vmin$.
The inequality given in Eq. (A.4) of \cite{Fox:2014kua}, ${\rm d}\tilde\eta/{\rm d}E_R$ in our notation, is actually an infinite set of inequalities, one for each $E_R$.

The KKT conditions have been extended to functionals defined on specific kinds of function spaces and constraints.
Banach spaces have been considered extensively (see e.g. \cite{Burger}). However, the functions we are looking for, i.e. step functions, do not have derivatives everywhere unless interpreted as distributions, and the spaces of distributions defined on non-compact intervals like $[0,\infty)$ are not Banach (under the usual weak-* topology).
More general spaces, i.e. locally convex topological vector spaces, have been considered by Dubovitskii and Milyutin \cite{D-M}.
As explained in the book by R. B. Holmes \cite{Holmes} (see pages 51 to 53), the Dubovitskii and Milyutin theory applies to constrained sets of functions that have non-empty interiors. However, the set of non-increasing functions such as $\tilde\eta$ has empty interior.
A function $\tilde\eta\in S$ is in the interior of a set $S$ if there is a neighborhood around it that belongs to the set, but for non-increasing functions there is no such neighborhood.
This is because a non-increasing function always has a non-monotonic function arbitrarily close to it.

Since we did not find in the literature any proof that clearly applies to our problems, in the following we present our own proof by first discretizing the variable $\vmin$ into a finite set of values $\vmin^i$ so that the KKT conditions are applicable, and then taking the continuum limit at the end.
Our proof is heuristic only in that it does not address the convergence of the limit.

For convenience, let us define a different functional of $\tilde\eta$, ($-2$ times the log-likelihood),
\begin{equation}
L[\tilde\eta]=-2\ln{\cal L}[\tilde\eta].
\label{eq:defL}
\end{equation}
With this definition, finding the $\tilde\eta$ function that maximizes the extended likelihood is equivalent to finding the function that minimizes $L$.
To simplify the problem, we discretize the $\vmin$ space into a set of $K+1$ positive variables $\vmin^i=v_\delta +i\times \Delta v$ with $i=0,1,\dots,K$, where $\Delta v\equiv (v^{\rm MAX}-v_\delta)/K$ with a large enough constant $v^{\rm MAX}$ value.
At the end, we will take $K\rightarrow\infty$ while keeping $v^{\rm MAX}$ constant.

With a $K$-dimensional vector $\vec{\tilde\eta}=(\tilde\eta_0,\tilde\eta_1,\dots,\tilde\eta_{K-1})$, we can define a piecewise constant function $\tilde\eta(v_{\rm min};\vec{\tilde\eta})$ given by
\begin{equation}
\tilde\eta(\vmin;\vec{\tilde\eta})\equiv \tilde\eta_i
\mbox{ if } \vmin^i\leq \vmin< \vmin^{i+1}.
\label{eq:discreteEta}
\end{equation}
Notice that there is no loss of generality of the $\tilde\eta(\vmin)$ considered, since any physically meaningful function is the limit of a sequence of piecewise constant functions as the number of steps tends to infinity.
The corresponding $L$ functional becomes a function $f_L$ of the vector $\vec{\tilde\eta}$, 
\begin{equation}
f_L(\vec{\tilde \eta})\equiv L[\tilde\eta(\vmin;\vec{\tilde\eta})].
\label{eq:functionL}
\end{equation}
With this discretization we can formalize the minimization of the functional $L$ as a limit of the function minimization of $f_L$, and by doing so we can safely apply the KKT conditions.

The KKT conditions for minimizing the function $f_L(\vec{\tilde\eta})$ under the constraints $\tilde\eta_{i}\ge \tilde\eta_{i+1}$ on its variables (i.e.~requiring the piecewise constant function $\tilde\eta(v_{\rm min};\vec{\tilde\eta})$ to be non-increasing) are the minimization conditions for the function
\begin{equation}
f_L'(\vec{\tilde\eta},\vec q)\equiv f_L(\vec{\tilde\eta})+\sum_{\i=0}^{K-1}q_i(\tilde\eta_{i+1}-\tilde\eta_i)
\label{eq:fL_prime}
\end{equation}
with respect to the variables $\vec{\tilde\eta}$ and $\vec q\equiv(q_0,\dots,q_{K-1})$ considered as unconstrained, and the supplementary conditions $q_i\geq0$ and $q_i(\tilde\eta_{i+1}-\tilde\eta_i)=0$.
Written explicitly, the KKT conditions are:
\begin{eqnarray}
(\rm KKT~I)_a&\quad& \frac{\partial f_L}{\partial \tilde\eta_i}+q_{i-1}-q_i=0\mbox{, for } 1\leq i\leq K-1,
  \label{eq:KKT-Ia}\\
(\rm KKT~I)_b&\quad& \frac{\partial f_L}{\partial \tilde\eta_0}-q_0=0,
  \label{eq:KKT-Ib}\\
(\rm KKT~II)&\quad& q_i\geq 0,
  \label{eq:KKT-II}\\
(\rm KKT~III)&\quad& \tilde \eta_{i+1}-\tilde \eta_{i}\leq 0, \mbox{ and}
  \label{eq:KKT-III}\\
  (\rm KKT~IV)&\quad& q_i(\tilde\eta_{i+1}-\tilde\eta_i)=0 \mbox{, or equivalently, }\nonumber\\
  && q_i(\tilde\eta_{i+1}-\tilde\eta_i)/\Delta v =0
  \mbox{ (no summation imposed).}
  \label{eq:KKT-IV}
\end{eqnarray}

Choosing $\vec{\tilde\eta}$ to be a unit vector $\hat{\tilde\eta}_i$ along the $i$th component $\tilde\eta_i$, the first term on the left-hand side of \eqref{eq:KKT-Ia} and \eqref{eq:KKT-Ib} can be written as
\begin{eqnarray}
\frac{\partial}{\partial\tilde\eta_i}f_L(\vec{\tilde\eta})
&=&\hat{\tilde\eta}_i\cdot\frac{\partial}{\partial\vec{\tilde\eta}}f_L(\vec{\tilde\eta})
=\lim_{\epsilon\rightarrow0}\frac{f_L(\vec{\tilde\eta}+\epsilon\hat{\tilde\eta}_i)-f_L(\vec{\tilde\eta})}{\epsilon}.
  \label{eq:partialfL}
\end{eqnarray}
Using \eqref{eq:partialfL}, we now have
\begin{eqnarray}
\frac{\partial}{\partial\tilde\eta_i}f_L(\vec{\tilde\eta})
&=&\lim_{\epsilon\rightarrow0}\frac{L[\tilde\eta(v_{\rm min};\vec{\tilde\eta}+\epsilon\hat{\tilde\eta}_i)]-L[\tilde\eta(v_{\rm min};\vec{\tilde\eta})]}{\epsilon}.
  \label{eq:partialfL2}
\end{eqnarray}
Using $L[\tilde\eta(\vmin;\vec{\tilde\eta}+\epsilon\hat{\tilde\eta}_i)]=L[\tilde\eta(\vmin;\vec{\tilde\eta})+\epsilon\tilde\eta(\vmin;\hat{\tilde\eta}_i)]$, \eqref{eq:partialfL2} can be written in terms of the functional derivative of the $L$ functional,
\begin{eqnarray}
\frac{\partial}{\partial\tilde\eta_i}f_L(\vec{\tilde\eta})
&=&\lim_{\epsilon\rightarrow0}\frac{L[\tilde\eta(\vmin;\vec{\tilde\eta})+\epsilon\tilde\eta(\vmin;\hat{\tilde\eta}_i)]-L[\tilde\eta(\vmin;\vec{\tilde\eta})]}{\epsilon}\\
&=&\int_0^\infty{\rm d}\vmin~\tilde\eta(\vmin;\hat{\tilde\eta}_i)\frac{\delta L}{\delta\tilde\eta(\vmin)}.
\label{eq:partialfL3}
\end{eqnarray}
From \eqref{eq:discreteEta} one can easily see that the function $\tilde\eta(\vmin;\hat{\tilde\eta}_i)$ in \eqref{eq:partialfL3} has a rectangular shape with value $1$ between $v^i_{\rm min}$ and $v^{i+1}_{\rm min}$ and zero everywhere else.
The summation over $i$ from $i=0$ to $j\leq K-1$ of the left-hand side of \eqref{eq:KKT-Ia} and \eqref{eq:KKT-Ib} is thus
\begin{eqnarray}
\frac{\partial f_L}{\partial \tilde\eta_0}-q_0+\sum_{i=1}^j\left(\frac{\partial f_L}{\partial \tilde\eta_i}+q_{i-1}-q_i\right)&=&\sum_{i=0}^j\frac{\partial f_L}{\partial\tilde\eta_i}-q_j\\
&=&\sum_{i=0}^j\int_0^\infty{\rm d}\vmin~\tilde\eta(\vmin;\hat{\tilde\eta}_i)\frac{\delta L}{\delta\tilde\eta(\vmin)}-q_j.
\label{eq:sumKKT-I}
\end{eqnarray}
Note that in the integrand \eqref{eq:sumKKT-I} $\sum_{i=0}^j\tilde\eta(\vmin;\hat{\tilde\eta}_i)=\theta(\vmin^j-\vmin)\theta(\vmin-v_\delta)$, thus
\begin{eqnarray}
  \sum_{i=0}^j\int_0^\infty{\rm d}\vmin~\tilde\eta(\vmin;\hat{\tilde\eta}_i)\frac{\delta L}{\delta\tilde\eta(\vmin)}-q_j=\int_{v_\delta}^{\vmin^j}{\rm d}\vmin~\frac{\delta L}{\delta\tilde\eta(\vmin)}-q_j.
  \label{eq:sumKKT-I2}
\end{eqnarray}
Using an interpolation function $q(\vmin)$ satisfying $q(\vmin^j)=q_j$, we finally conclude, using \eqref{eq:sumKKT-I2}, that the ${\rm KKT~I}$ conditions \eqref{eq:KKT-Ia} and \eqref{eq:KKT-Ib} imply
\begin{equation}
  \int_{v_\delta}^{\vmin^j}{\rm d}v~\frac{\delta L}{\delta\tilde\eta(v)}-q(\vmin^j)=0.
  \label{eq:intL}
\end{equation}

By taking a large enough $K$ with $v^{\rm MAX}$ fixed, we can find an integer $j$ such that $v^j_{\rm min}$ is arbitrarily close to a given $\vmin$ value, if $v_{\delta}\leq v_{\rm min}\leq v^{\rm MAX}$.
Therefore, in the limit $K\rightarrow\infty$, and thus $\Delta v\rightarrow 0$, we can write the conditions \eqref{eq:intL} and \eqref{eq:KKT-II} to \eqref{eq:KKT-IV} for continuous $v_{\rm min}$ and $\tilde\eta$ variables:
\begin{eqnarray}
  ({\rm I})&\quad& q(\vmin)= \int_{v_\delta}^{\vmin}{\rm d}v~\frac{\delta L}{\delta \tilde\eta(v)},
  \label{eq:I}\\
({\rm II})&\quad& q(\vmin)\geq 0,\label{eq:II}\\
({\rm III})&\quad& \forall \epsilon>0,\quad \tilde \eta (\vmin+\epsilon)\leq \tilde \eta (\vmin), \mbox{ and}\label{eq:III}\\
({\rm IV})&\quad& q(\vmin)\lim_{\epsilon\rightarrow+0}\dfrac{\tilde\eta(\vmin+\epsilon)-\tilde\eta(\vmin)}{\epsilon}=0.\label{eq:IV}
\end{eqnarray}
Note that although we write the conditions in terms of continuous variables, they should always be understood as a limit of the conditions for discrete variables.

Two direct consequences of $({\rm IV})$ \eqref{eq:IV} are: i) $\tilde\eta(\vmin)$ can be discontinuous only at the points where $q(\vmin)$ vanishes, and ii) $\tilde\eta(\vmin)$ is constant in an open interval where $q(\vmin)\neq0$.
If there is an open interval where $q(\vmin)$ is zero, within the interval, $({\rm IV})$ is trivially satisfied.
Therefore, $\tilde\eta(\vmin)$ is a piecewise constant function with discontinuity points where $q(\vmin)=0$.
Let us examine the possible zeros of the $q(\vmin)$ function.

Using \eqref{eq:extendedLikelihood} and \eqref{eq:defL} in $({\rm I})$ \eqref{eq:I}, we get 
\begin{eqnarray}\label{eq:qofvmin_first}
q(\vmin)&=& 2\int_{v_\delta}^{\vmin}{\rm d}v~\frac{\delta N_E}{\delta \tilde\eta(v)}
-2\int_{v_\delta}^{\vmin}{\rm d}v~\sum_{i=a}^{N_O}\frac{\delta}{\delta \tilde\eta(v)}\ln\left(\left.\frac{{\rm d}R_{tot}}{{\rm d}E'}\right|_{E'=E'_a}\right).
\end{eqnarray}
In \eqref{eq:NE} $N_E$ is given in terms of ${\cal R}$, given in turn in \eqref{eq:energy_integr_resp}, where $d{\cal R}/dE'$ is in \eqref{eq:calR}.
Using these equations, \eqref{eq:qofvmin_first} becomes
\begin{eqnarray} \label{eq:qofvmin_second}
q(\vmin)&=& 2 MT\int_{E'_{\rm min}}^{E'_{\rm max}}{\rm d}E'~\frac{{\rm d}{\cal H}}{{\rm d}E'}(E', \vmin) \nonumber\\
&&\qquad\qquad\qquad
-2\sum_{a=1}^{N_O}\left.\left[\int_{v_\delta}^{\vmin}{\rm d}v~\left(\frac{\delta}{\delta \tilde\eta(v)}\frac{{\rm d}R_{tot}}{{\rm d}E'}\right)\right]_{E'=E'_a}\middle/\left[\frac{{\rm d}R_{tot}}{{\rm d}E'}\right]_{E'=E'_a}\right..
\end{eqnarray}
We define:
\begin{equation}
\xi(\vmin)
\equiv MT\int_{E'_{\rm min}}^{E'_{\rm max}}{\rm d}E'~\frac{{\rm d}{\cal H}}{{\rm d}E'}(E', \vmin).
\label{eq:defxi}
\end{equation}
Using \eqref{eq:dRtot_dE} and \eqref{eq:calR}, we can write
\begin{eqnarray}
\int_{v_\delta}^{\vmin}{\rm d}v&&\left.\left(\frac{\delta}{\delta \tilde\eta(v)}\frac{{\rm d}R_{tot}}{{\rm d}E'}\right)\right|_{E'=E'_a}
= \int_{v_\delta}^{\vmin}{\rm d}v~\left.\left(\frac{{\rm d}{\cal R}}{{\rm d}E'}\right)\right|_{E'=E'_a} \nonumber \\
&&= \left.\frac{{\rm d}{\cal H}}{{\rm d}E'}(E',\vmin)\right|_{E'=E'_a} \equiv H_a(\vmin)
\label{eq:defHa}
\end{eqnarray}
and
\begin{equation}
\gamma_a[\tilde\eta] \equiv \left.\frac{{\rm d}R_{tot}}{{\rm d}E'}\right|_{E'=E'_a}.
\label{eq:defgammaa}
\end{equation}
Replacing \eqref{eq:defxi} to \eqref{eq:defgammaa} into \eqref{eq:qofvmin_second}, we obtain
\begin{eqnarray}
q(\vmin)&\equiv& 2\xi(\vmin)-2\sum_{a=1}^{N_O}\frac{H_a(\vmin)}{\gamma_a[\tilde\eta]}.
\label{eq:qv}
\end{eqnarray}
In this equation, the only $\tilde\eta$ dependence is in $\gamma_a[\tilde\eta]$.
The functions $\xi(\vmin)$ and $H_a(\vmin)$ do not depend on $\tilde\eta$.

\begin{figure}[t]
\centering
\includegraphics[width=0.48\textwidth]{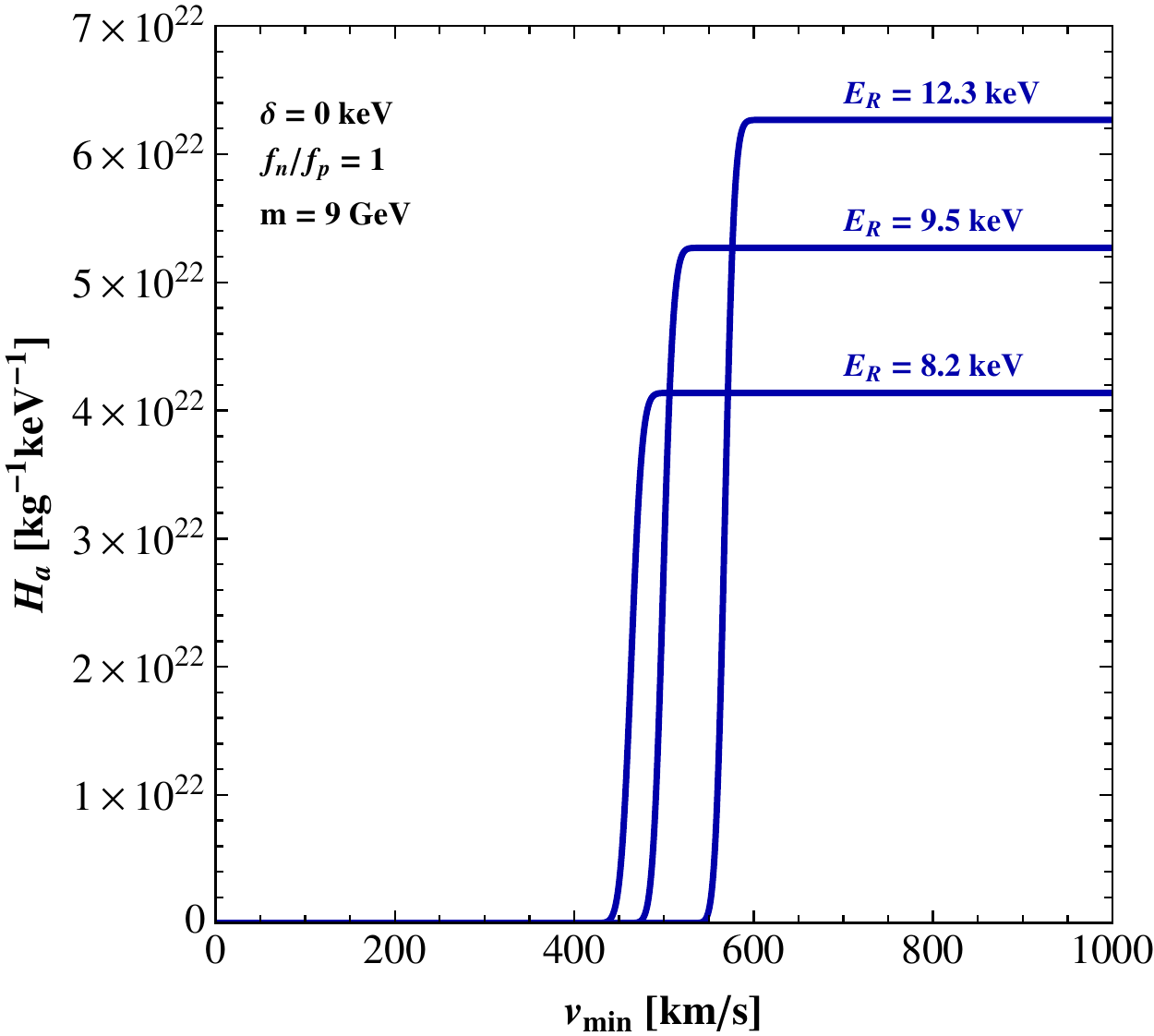}
\includegraphics[width=0.48\textwidth]{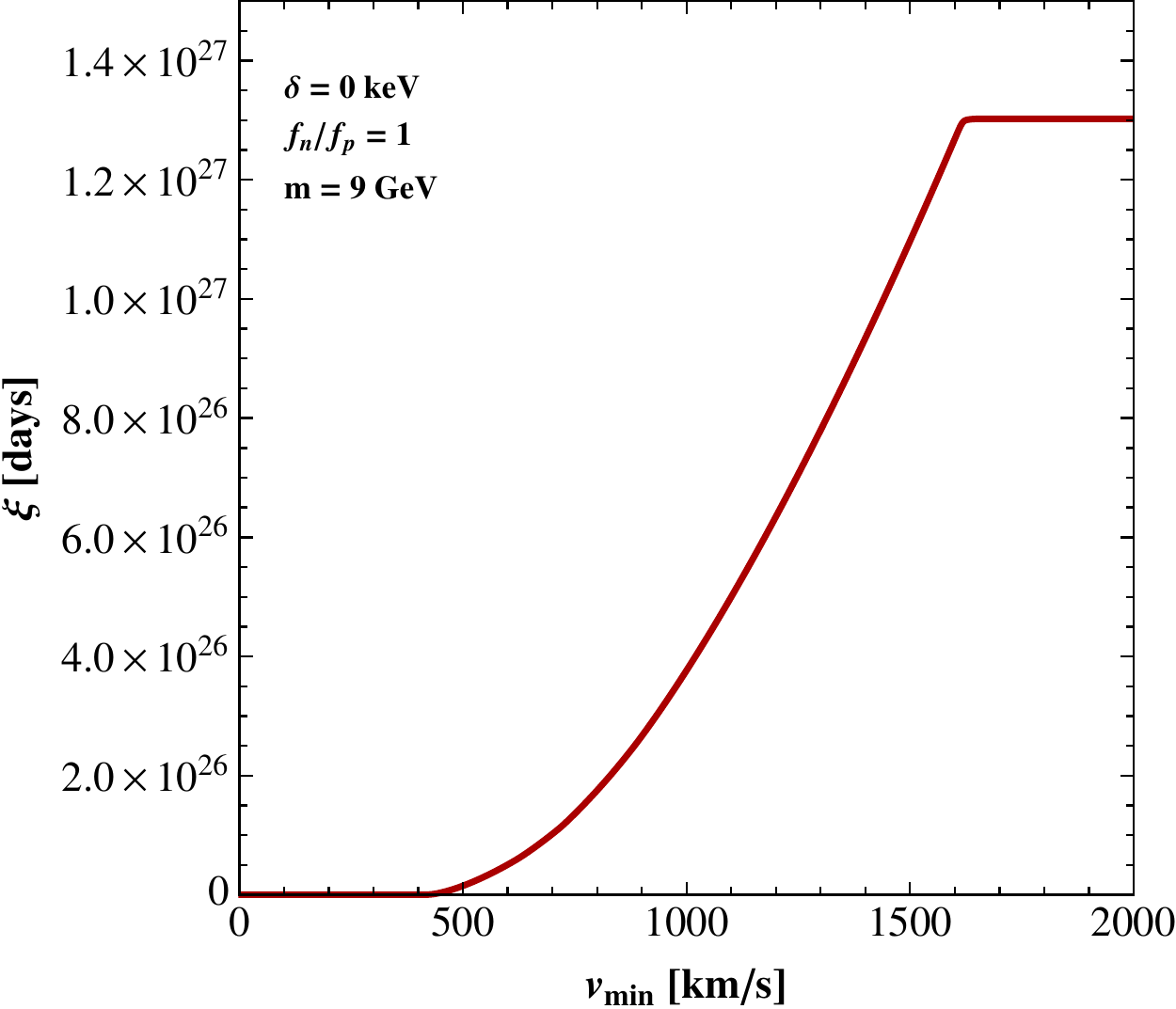}
\caption{$H_a(\vmin)$ (left panel) and $\xi(\vmin)$ (right panel) for elastic isospin-conserving SI interactions and $m=9$ GeV, for the three events of CDMS-II-Si.}
\label{fig:HaXi}
\end{figure}

Fig.~\ref{fig:HaXi} shows the functions $H_a(v_{\rm min})$ and $\xi(v_{\rm min})$ for the three candidate events of CDMS-II-Si assuming an SI cross section with $f_n/f_p=1$ and $m=9$ GeV.
In order to explain the form of these functions, let us first consider a simple situation where the target material consists of a single nuclide, or multiple isotopes of the same element, as in CDMS-II-Si.
In this case, the integrands of the different terms ${\rm d}{\cal H}_T/{\rm d}E'$ in \eqref{eq:calHT} contributing to ${\rm d}{\cal H}/{\rm d}E'$ in \eqref{eq:calH} are similarly localized in $E_R$ for all nuclides $T$ (for a fixed $E'$).
Notice that these integrands are independent of $v$ if $d\sigma_T/dE_R$ is proportional to $v^{-2}$.
In this case, the $v$ dependence of ${\rm d}{\cal H}_T/{\rm d}E'$ is only in the integration range $[E_R^{T-}(v),E_R^{T+}(v)]$.
If so, as $v$ increases, this range covers more of the region in which the integrand is non-zero.
Thus, ${\rm d}{\cal H}/{\rm d}E'$ grows with $v$ in a certain range.
When $v$ is large enough for the integration in \eqref{eq:calHT} to cover all the region in which the integrand is non-zero, ${\rm d}{\cal H}_T/{\rm d}E'$ becomes constant, and so does ${\rm d}{\cal H}/{\rm d}E'$.
This explains the step-like functional form of $H_a(\vmin)$ given in \eqref{eq:defHa}, which is ${\rm d}{\cal H}/{\rm d}E'$ with $E'=E_a$, as can be seen in the left panel of Fig.~\ref{fig:HaXi}.

Looking at \eqref{eq:calHT} which defines ${\rm d}{\cal H}_T/{\rm d}E'$ for each nuclide $T$, we see that the only dependence on $E'$ of the integrand is in $\epsilon(E_R, E')G_T(E_R, E')$.
To compute $\xi(\vmin)$, we need thus a double integration, first in $E_R$ to obtain ${\rm d}{\cal H}_T/{\rm d}E'$, and then in $E'$, after summing all ${\rm d}{\cal H}_T/{\rm d}E'$ contributing to ${\rm d}{\cal H}/{\rm d}E'$.
If we exchange the order of integration, performing the $E'$ integration first, we see that as $E_R$ increases, for $E_R$ very small the integrand $\epsilon G_T$ will be zero within the $E'$ integration range.
Then, the non-zero portion of $\epsilon G_T$ within the $E'$ integration range will increase, then be entirely contained, and then decrease and become zero again.
Thus, the resulting integrand in $E_R$ will be slowly changing in the $E_R$ range in which it is non-zero.
As $\vmin$ increases, the integration range in $E_R$ encompasses more of the slowly varying integrand, resulting in a smoothly increasing function $\xi(\vmin)$, as shown in the right panel of Fig.~\ref{fig:HaXi}.
Once $\vmin$ becomes large enough for the integration range in $E_R$ to cover all the non-zero part of the integrand in $\xi(\vmin)$, this function becomes constant (see Fig.~\ref{fig:HaXi}).

\begin{figure}[t]
\centering
\includegraphics[width=0.47\textwidth]{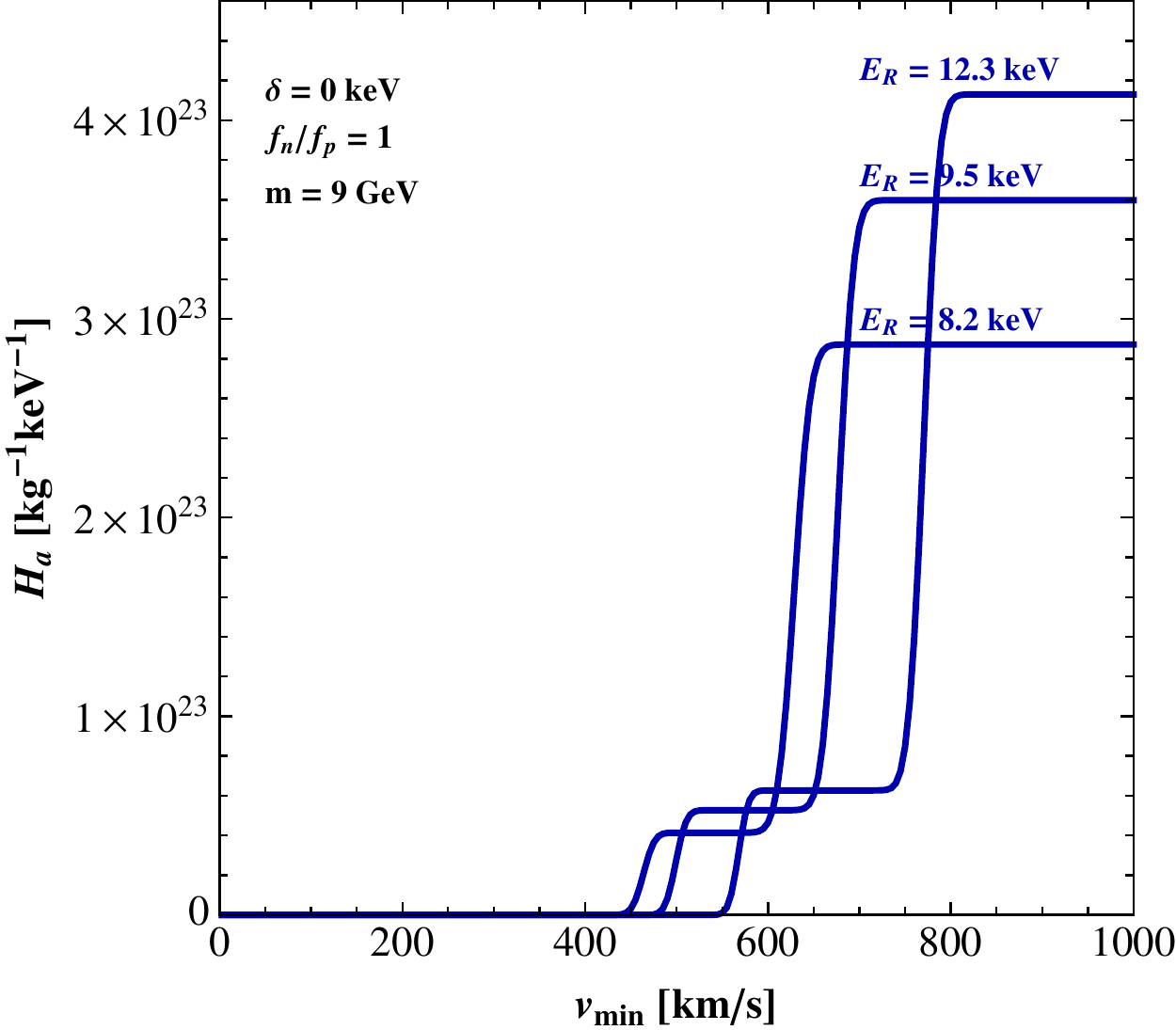}
\includegraphics[width=0.51\textwidth]{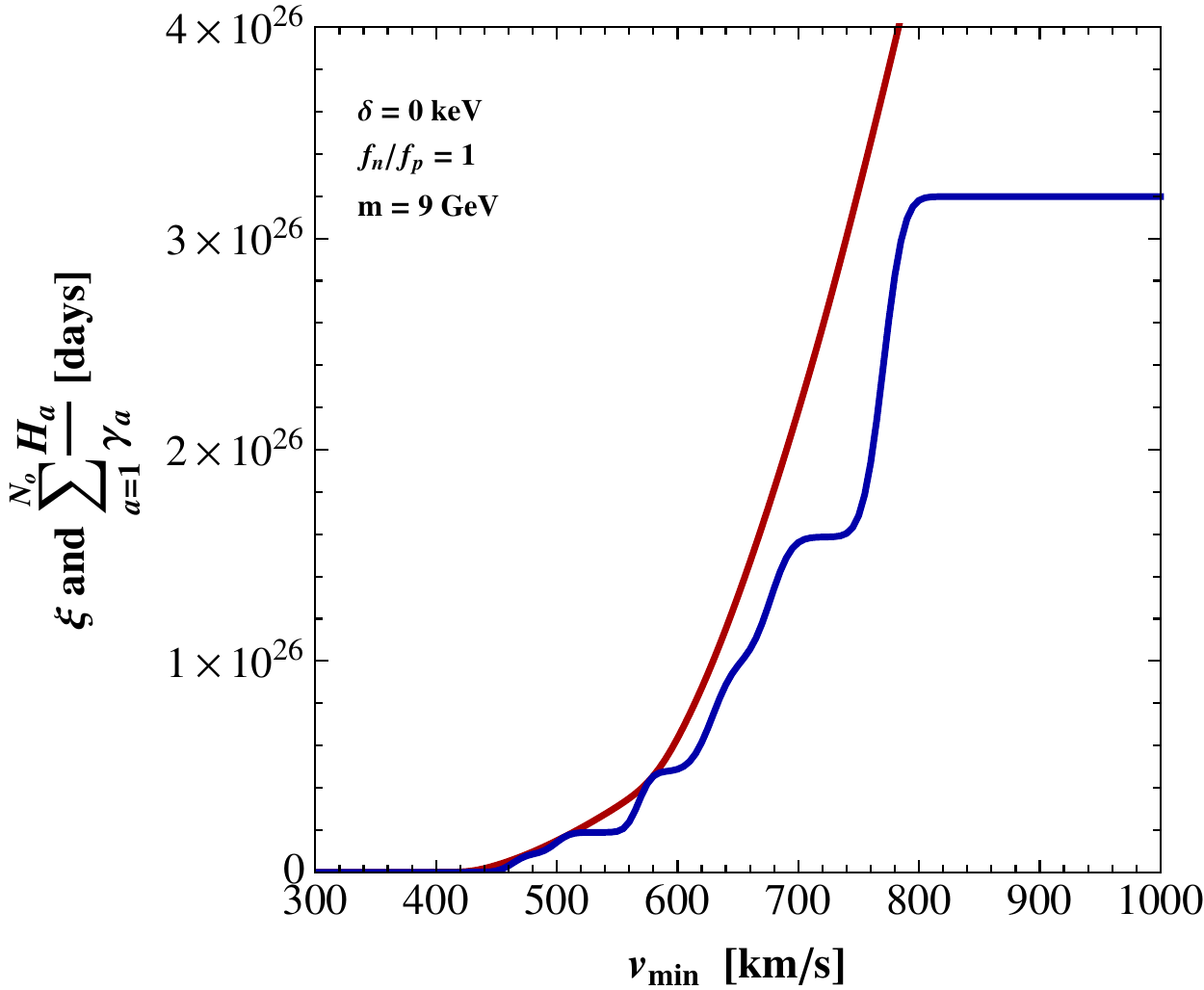}
\caption{Same as Fig.~\ref{fig:HaXi} but for a fictitious detector with target material composed of equal mass fractions of Si and Ge (see the text), and showing in addition $\sum_a H_a(\vmin)/\gamma_a$ (blue line, right panel) .}
\label{fig:qvSiGe}
\end{figure} 

Let us return to study the discontinuity points of the best-fit $\tilde\eta$ function which happen at the zeros of $q(\vmin)$ given in \eqref{eq:qv}.
For elastic ($\delta=0$) or endothermic ($\delta>0$) scattering, there is a region at small $\vmin$ values where both $H_a$ and $\xi$ vanish.
Looking at \eqref{eq:calHT}, when $E_R^{T+}(v)$ is below the experimental threshold, the integrand, in particular the acceptance $\epsilon$, is zero, thus ${\rm d}{\cal H}/dE'=0$.
In this $\vmin$ region the condition $q(\vmin)=0$ is trivially satisfied, and the shape of the best-fit $\tilde\eta$ function is undetermined.

Changes in $\tilde\eta(\vmin)$ produce changes in $\gamma_a[\tilde\eta]$.
For values of $\gamma_a$ which make the second term of the right hand side of (3.30) large enough to reach the first term $2\xi(\vmin)$ from below, $q(\vmin)$ (see \eqref{eq:qv}) has non-trivial zeros where $\xi$ and $H_a$ are non-zero.
The non-negativity of $q(v)\geq0$ means that $q(v)=0$ only when the monotonically increasing function $\xi$ touches the step-like $\sum_a H_a/\gamma_a$ function from above.
Since $\sum_a H_a/\gamma_a$ has $N_O$ steps, this can happen only at a number of $\vmin$ values smaller than or equal to $N_O$.
Examples of these functions $\xi$, $\sum_a H_a/\gamma_a$, and $q$ will be shown below in Figs.~\ref{fig:qvdelta0}, \ref{fig:qvdeltaneg50} and \ref{fig:qvdeltaneg200}.

To guess the generic shape of the $\xi(\vmin)$ and $H_a(\vmin)$ functions for differential cross sections whose WIMP speed $v$ dependence is different from $\propto v^{-2}$, let us assume the differential cross section for a given $E_R$ behaves as $v^{(n-2)}$ for large values of $v$.
One such example is that of WIMPs interacting with nuclei through a magnetic dipole moment, where $n=2$ at large $v$ (see (3.9) of \cite{DelNobile:2014eta}).
In this case, from \eqref{eq:calHT} one can easily see that the shapes of the functions $H_a(v)/v^n$ and $\xi(v)/v^n$ should be similar to those of $H_a(v)$ and $\xi(v)$ for a differential cross section proportional to $v^{-2}$.
Therefore, the argument given above can be used for $q(v)/v^n$, whose zeros are the same as those of $q(v)$, leading to the same conclusions.

When the target consists of several elements, each $H_a$ has multiple step-like features, one for each element.
This is illustrated in Fig.~\ref{fig:qvSiGe} for a fictitious CDMS-II-like detector composed of equal mass fractions of Si and Ge.
We see in the left panel of Fig.~\ref{fig:qvSiGe} that for each of the three elements there are two step-like features in $H_a$.
One may naively expect that because in this case there are $2N_O$ step-like features in $\sum_a H_a/\gamma_a$, the number of zeros of the function $q(\vmin)$ would equally double.
However, this is not the case.
Because $\xi$ and $H_a$ are independent of $\tilde\eta$, by changing $\tilde\eta$ and thus $\gamma_a$ in general one can make at most one of the two steps per observed event in $\sum_a H_a/\gamma_a$ touch the function $\xi(\vmin)$ from below.
Thus the number of zeros of $q(\vmin)$ is still at most $N_O$.
This can be seen in the right panel of Fig.~\ref{fig:qvSiGe}.

In summary, in this section we proved that the $\tilde\eta$ function maximizing the extended likelihood is a piecewise constant function with a number of steps smaller than or equal to the number $N_O$ of observed events.

\section{EHI analysis in the $\vmin$-space}\label{sec:EHI}

In this section we show how to find the solution to the maximization of the extended likelihood in the EHI method, in the $\vmin$--$\tilde\eta$ space.
As shown in the previous section, the best-fit function, which we call $\tilde\eta_{\rm BF}(\vmin)$ from now on, is a piecewise constant function with at most $N_O$ steps (note that in the statistics literature the subscript ``ML'' for maximum likelihood is usually used instead of ``BF'').
We will also find a statistically meaningful confidence band around $\tilde\eta_{\rm BF}(\vmin)$, which we will define as a pointwise confidence band.

\subsection{Finding the best-fit function $\tilde\eta_{\rm BF}(v_{\rm min})$}\label{subsec:BF}

The properties of the $\tilde\eta$ function maximizing the extended likelihood we have proven in the previous section can be utilized to find $\tilde\eta_{\rm BF}$.
We can define a function $f_L^{(N_O)}$ of $2N_O$ variables, $\vec{v}=(v_1,v_2,\dots,v_{N_O})$ and $\vec{\tilde\eta}=(\tilde\eta_1,\tilde\eta_2,\dots,\tilde\eta_{N_O})$, specifying the positions and heights of the $N_O$ steps, as a restriction of the functional $L[\tilde\eta]$:
\begin{equation}
f_L^{(N_O)}(\vec{v},\vec{\tilde\eta})
\equiv
L[\tilde\eta^{(N_O)}(\vmin;\vec{v},\vec{\tilde\eta})].
\label{eq:fL_NO}
\end{equation}
The piecewise constant function $\tilde\eta^{(N_O)}$ is defined as
\begin{eqnarray}
\tilde\eta^{(N_O)}(\vmin;\vec{v},\vec{\tilde\eta})
\equiv
  \begin{cases}
    \tilde\eta_a&\text{if}~ v_{a-1} < \vmin \leq v_a,\\
    0&\text{if}~ v_{N_O} < \vmin,
  \end{cases}
 \label{eq:etaNO}
\end{eqnarray}
where $a=1, \dots, N_O$.
Here we assume $\vmin$ and $v_a$'s are all larger than $v_\delta$, and the constraints \eqref{eq:III} $\tilde\eta_a\leq\tilde\eta_b$ for $a>b$ are satisfied.
Since the function $\tilde\eta$ cannot change after the last step and it must reach zero for large $\vmin$, it must be zero for $\vmin>v_{N_O}$.
We do not specify the value of $\tilde\eta^{(N_O)}$ below the minimum $v_\delta$ since the event rate is independent of it.
Notice that \eqref{eq:etaNO} requires the definition of $v_0$.
We define $v_0=v_\delta$ for convenience.

From these definitions and the theorem we have proven, we can easily obtain $\tilde\eta_{\rm BF}$ and $L_{\rm min}$, the minimum value of the functional $L[\tilde\eta]$, by finding $\vec v_{\rm BF}$ and $\vec{\tilde\eta}_{\rm BF}$ that minimize $f_L^{(N_O)}$, so that
\begin{equation}
\tilde\eta_{\rm BF}(\vmin)=\tilde\eta^{(N_O)}(\vmin;\vec v_{\rm BF},\vec{\tilde\eta}_{\rm BF})
\end{equation}
and
\begin{eqnarray}
L_{\rm min}&\equiv&L[\tilde\eta_{\rm BF}(\vmin)]=L[\tilde\eta^{(N_O)}(\vmin;\vec v_{\rm BF},\vec{\tilde\eta}_{\rm BF})].
\end{eqnarray}

From the definition \eqref{eq:extendedLikelihood} of the extended likelihood function, we can write $f_L^{(N_O)}$ in a simple form as
\begin{eqnarray}
f_L^{(N_O)}
&=&2N_{BG}+2MT\sum_{a=1}^{N_O}\tilde\eta_a\int_{v_{a-1}}^{v_a}{\rm d}\vmin
{\cal R}_{[E'_{\rm min},E'_{\rm max}]}(\vmin)\nonumber\\
&&-2\sum_{i=1}^{N_O}\ln\left[MT\sum_{a=1}^{N_O}\tilde\eta_a
\int_{v_{a-1}}^{v_a}{\rm d}\vmin
\frac{{\rm d}{\cal R}}{{\rm d}E'}(\vmin)
+MT\frac{dR_{\rm BG}}{{\rm d}E'}(\vmin)\right]_{E'=E'_i},
\label{eq:flNO}
\end{eqnarray}
with $N_\text{BG}$ given in \eqref{eq:NBG}.
Defining the ${\cal N}_a$ and ${\cal M}_{ai}$ functions of $v_a$ as
\begin{eqnarray}
{\cal N}_a(\vec v)&\equiv&
MT\int_{v_{a-1}}^{v_a}{\rm d}\vmin
{\cal R}_{[E'_{\rm min},E'_{\rm max}]}(\vmin),\\
{\cal M}_{ai}(\vec v)
&\equiv&MT\int_{v_{a-1}}^{v_a}{\rm d}\vmin
~\left.\frac{{\rm d}{\cal R}}{{\rm d}E'}\right|_{E'=E'_i}(\vmin),
\end{eqnarray}
and the fixed constants $b_i$
\begin{eqnarray}
b_i&\equiv& MT \left.\frac{{\rm d}{R_{\rm BG}}}{{\rm d}E'}\right|_{E'=E'_i},
\end{eqnarray}
we can write \eqref{eq:flNO} as
\begin{eqnarray}
f_L^{(N_O)}&=&2N_\text{BG} + 2\sum_{a=1}^{N_O}\tilde\eta_a {\cal N}_a
-2\sum_{i=1}^{N_O}\ln\left[\sum_{a=1}^{N_O}\tilde\eta_a {\cal M}_{ai}+b_i\right].
\label{eq:flNOsum}
\end{eqnarray}

The minimization of the function $f_L^{(N_O)}$ of $2N_O$ parameters $v_1,\dots,v_{N_O},\:\tilde\eta_1,\dots,\tilde\eta_{N_O}$, subject to the constraints 
\begin{eqnarray}
 &v_1&> v_\delta, 
 \label{eq:constr_v_positive} \\
 &v_b&-v_a \ge 0 \text{ and }\: \tilde\eta_a - \tilde\eta_b \ge 0 \:\text{ for } a<b,
 \label{eq:constr_steps}
\end{eqnarray}
can be done numerically using a global minimization algorithm.
In the implementation, we express $f_L^{(N_O)}$ in terms of $\ln \tilde\eta_a$ and use $\ln \tilde\eta_a$ instead of $\tilde\eta_a$ as variables, since $\tilde\eta_a$ span many orders of magnitude.
This also accounts for the $\tilde\eta_a > 0$ constraints, leaving only the constraints in \eqref{eq:constr_v_positive} and \eqref{eq:constr_steps} to be enforced in the minimization.
Note that in general minimization algorithms may attempt to evaluate the function in regions where the constraints are not satisfied, and in these regions the function $f_L^{(N_O)}$ is not well defined, thus a fictitious function must be used that grows smoothly with the absolute value of the unsatisfied constraints in \eqref{eq:constr_v_positive} and \eqref{eq:constr_steps}.

\subsection{Finding the confidence band}

In order to compare the $\tilde\eta_\text{BF}$ we obtained with the upper limits imposed by other experiments, we need a way to represent the uncertainty in our determination of $\tilde\eta_{\rm BF}$.
This can be achieved by finding a region in the $\vmin$--$\tilde\eta$ space satisfying a certain statistical criterion, analogous to the confidence interval in the usual analysis with a fixed halo model.
The region in the $\vmin$--$\tilde\eta$ space which is densely filled by the family of all possible $\tilde\eta(\vmin)$ curves satisfying 
\begin{equation}
\Delta L[\tilde\eta]\equiv L[\tilde\eta]-L_{\rm min}\leq \Delta L^*,
\label{eq:deltaL}
\end{equation}
with given $\Delta L^*$, is a natural candidate to examine.
The condition in \eqref{eq:deltaL} defines a two-sided interval around $\tilde\eta_{\rm BF}$ for each $\vmin$ value, and the collection of those intervals forms a pointwise confidence band in $\vmin$--$\tilde\eta$ space.
From now on we will call it simply ``the confidence band''.

Conceptually, computing the confidence band is a straightforward procedure, but in practice, finding all the $\tilde\eta$ functions satisfying \eqref{eq:deltaL} and constructing the band from them is not possible.
If the same band can be formed by a much smaller subset of them, and this subset is much easier to find than the whole set, the construction of the band would be practical.

As a possible subset, let us consider the set of $\tilde\eta$ functions which minimize $L[\tilde\eta]$ subject to the constraint 
\begin{equation}
\tilde\eta(v^*)=\tilde\eta^*.
\label{eq:eta_constraint}
\end{equation}
Let us define $L^c_{\rm min}(v^*,\tilde\eta^*)$ to be the minimum of the $L[\tilde\eta]$ subject to the constraint \eqref{eq:eta_constraint}, and 
\begin{equation}
\Delta L^c_{\rm min}(v^*,\tilde\eta^*) = L^c_{\rm min}(v^*,\tilde\eta^*) - L_{\rm min}.
\label{eq:deltaL_c_min}
\end{equation}
If $\Delta L^c_{\rm min}(v^*,\tilde\eta^*)$ is larger than a chosen $\Delta L^*$, it simply means that the point $(v^*,\tilde\eta^*)$ lies outside of the confidence band.
If it were inside the band, there should be at least one $\tilde\eta$ function passing through the $(v^*, \tilde\eta^*)$ point, for which $\Delta L[\tilde\eta]\le\Delta L^*$, in contradiction with the fact that $\Delta L^c_{\rm min}(v^*,\tilde\eta^*)>\Delta L^*$.
On the other hand, if $\Delta L^c_{\rm min}(v^*,\tilde\eta^*)\le\Delta L^*$, the confidence band should cover the point $(v^*,\tilde\eta^*)$ by definition.
Therefore, by finding the range of $\tilde\eta^*$ values which satisfy $\Delta L^c_{\rm min}(v^*,\tilde\eta^*)\leq\Delta L^*$ for each $v^*$ value, we can construct the band.

The remaining problem is how to find an easy way of computing $L^c_{\rm min}(v^*,\tilde\eta^*)$ (and therefore $\Delta L^c_{\rm min}(v^*,\tilde\eta^*)$).
We will now prove that the $\tilde\eta$ function minimizing $L[\tilde\eta]$ subject to the constraint \eqref{eq:eta_constraint} should be a piecewise constant function with at most $N_O+1$ discontinuities.

Let us rewrite the KKT conditions in \eqref{eq:KKT-Ia}-\eqref{eq:KKT-IV} but now with an additional equality constraint 
\begin{equation}
\tilde\eta_k = \tilde\eta^*,
\label{eq:eta_k_constraint}
\end{equation}
where the index $k$ is chosen to satisfy $v^k_{\rm min}\leq v^* <v^{k+1}_{\rm min}$, so that $\vmin^k$ can be arbitrarily close to $v^*$ for large enough $K$ values.
The additional constraint leads to the necessity of adding the term $p^*(\tilde\eta_k-\tilde\eta^*)$ to the function $f_L'$ in \eqref{eq:fL_prime} introducing a Lagrange multiplier $p^*$, so we define another function $f_L''(\vec{\tilde\eta},\vec q,p^*)$ as 
\begin{eqnarray}
f_L''(\vec{\tilde\eta},\vec q,p^*)
&\equiv& f_L'(\vec{\tilde\eta},\vec q)+ p^*(\tilde\eta_k-\tilde\eta^*)\\
&=& L[\tilde\eta(\vmin;\vec{\tilde\eta})]
+\sum_{\i=0}^{K-1}q_i(\tilde\eta_{i+1}-\tilde\eta_i)
+p^*(\tilde\eta_k-\tilde\eta^*),
\end{eqnarray}
and use it to derive new KKT conditions.

The new KKT conditions consist of the unconstrained minimization conditions of the function $f_L''(\vec{\tilde\eta},\vec q,p^*)$ with respect to the parameters $\vec{\tilde\eta}$, $\vec q$ and $p^*$, plus the complementary conditions, which are the same as before.
Therefore, besides the constraint \eqref{eq:eta_k_constraint}, the changes only appear in the $(\rm KKT~I)_a$ and $(\rm KKT~I)_b$ conditions, where the function $f'_L$ was present.
The new conditions are
\begin{eqnarray}
  (\rm KKT~I)'_a&\quad& \frac{\partial f_L}{\partial \tilde\eta_i}+q_{i-1}-q_i+p^*\delta_{ki}=0
  \mbox{, for } 1\leq i\leq K-1 \mbox{, and}
  \label{eq:newKKT-Ia}\\
  (\rm KKT~I)'_b&\quad& \frac{\partial f_L}{\partial \tilde\eta_0}-q_0+p^*\delta_{k0}=0
  \label{eq:newKKT-Ib}
\end{eqnarray}
with additional terms $p^*\delta_{ki}$ and  $p^*\delta_{k0}$, respectively, and the constraint \eqref{eq:eta_k_constraint}.
Following similar steps as those in Sec.~\ref{sec:ubHI} from \eqref{eq:partialfL} to \eqref{eq:intL}, the summation of \eqref{eq:newKKT-Ia} and \eqref{eq:newKKT-Ib} over $i$ from $0$ to $j$ now becomes
\begin{equation}
  \int_{v_\delta}^{\vmin^j}{\rm d}v~\frac{\delta L}{\delta\tilde\eta(v)}-q(\vmin^j)
  + p^*\theta(\vmin^j-v^k)
  =0.
\end{equation}
In the limit of $K\rightarrow\infty$, the first condition for the $\tilde\eta$ functions minimizing $L[\tilde\eta]$ subject to the constraint \eqref{eq:eta_constraint} becomes
\begin{eqnarray}
({\rm I})'&\quad& q(\vmin)= \int_{v_\delta}^{\vmin}{\rm d}v~\frac{\delta L}{\delta \tilde\eta(v)}  + p^*\theta(\vmin-v^*),
  \label{eq:new-I}
\end{eqnarray}
while the conditions $(\rm{II})$, $(\rm{III})$ and $(\rm{IV})$ are the same as in \eqref{eq:II}-\eqref{eq:IV}.

Using the definition of $L[\tilde\eta]$ in \eqref{eq:defL} and \eqref{eq:extendedLikelihood} in the condition \eqref{eq:new-I}, we can write the function $q(\vmin)$ as
\begin{equation}
q(\vmin) = 2\xi(\vmin)-2\sum_{a=1}^{N_O}\frac{H_a(\vmin)}{\gamma_a[\tilde\eta]}+p^*\theta(\vmin-v^*),
\label{eq:newqv}
\end{equation}
with $\xi(\vmin)$, $H_a(\vmin)$ and $\gamma_a[\tilde\eta]$ defined in \eqref{eq:defxi}, \eqref{eq:defHa} and \eqref{eq:defgammaa}, respectively.

Again, the conditions in \eqref{eq:new-I} and \eqref{eq:IV} tell that the $\tilde\eta$ function we find is piecewise constant with discontinuities only at the isolated zeros of $q(\vmin)$.
We already argued that $\xi(\vmin)$ can touch the function $\sum_{a=1}^{N_O} H_a/\gamma_a$ from above at a number of points equal to or less than the number of observed events $N_O$.
Since $p^*\theta(\vmin-v^*)$ introduces another step on the right hand side of \eqref{eq:newqv}, with the right $p^*$ value $q(\vmin)$ could have an additional zero.
Thus the $\tilde\eta(\vmin)$ function minimizing $L[\tilde\eta]$ subject to the constraint \eqref{eq:eta_constraint} is piecewise constant with at most $N_O+1$ discontinuities.

Using a function $\tilde\eta$ of this type in \eqref{eq:deltaL} for each $(v^*, \eta^*)$, we minimize $L[\tilde\eta]$ in \eqref{eq:defL} as in Sec.~\ref{subsec:BF} to compute $\Delta L^c_{\rm min}(v^*,\tilde\eta^*)$ in \eqref{eq:deltaL_c_min}.
We define a function $f_L^{(N_O+1)}(\vec{v},\vec{\tilde\eta})$ as in \eqref{eq:fL_NO}, parametrized by $\vec{v}=(v_1, v_2, \dots, v_i=v^*, \dots, v_{N_O+1})$ and $\vec{\tilde\eta}=(\tilde\eta_1, \tilde\eta_2, \dots, \tilde\eta_i=\tilde\eta^*, \dots, \tilde\eta_{N_O+1})$.
The minimization of $f_L^{(N_O+1)}$ can again be done numerically using a global minimization algorithm, subject to the same constraints as in \eqref{eq:constr_v_positive}-\eqref{eq:constr_steps}, where in addition we keep $(v_i, \tilde\eta_i)$ fixed at $(v^*, \tilde\eta^*)$.
As before, in our implementation of the algorithm we write  $f_L^{(N_O+1)}$ in terms of $\ln \tilde\eta_a$ instead of $\tilde\eta_a$.
We repeat the minimization procedure for all indices $i=1\dots (N_O+1)$ corresponding to the position of the $(v^*, \tilde\eta^*)$ step in $\tilde\eta$, and select the solution that gives the overall minimum of $f_L^{(N_O+1)}$.

\subsection{Statistical interpretation of the confidence band}

From the procedure described above we can get both the best-fit $\tilde\eta$ function, $\tilde\eta_\text{BF}(\vmin)$, and the confidence band.
For a quantitative assessment of the compatibility with other experimental data, we need to know the statistical meaning of a particular choice for $\Delta L^*$.
One may be tempted to interpret $\Delta L$ as $-2$ times the logarithm of the likelihood ratio with $2N_O$ parameters, since we parametrized the $\tilde\eta$ function with $2N_O$ parameters (plus $v^*$ and $\tilde\eta^*$ which are fixed each time) to obtain the confidence band.
However, this is not the proper interpretation.
Note that the defining properties of the best-fit $\tilde\eta_{\rm BF}$ and the band do not rely on how we compute them.

Let us return to the definition of $\Delta L^c_{\rm min}(v^*,\tilde\eta)$ and use again the discretization procedure introduced to derive the KKT conditions in Sec.~\ref{sec:ubHI}.
With a discretization of $\vmin$ we can define a likelihood function 
\begin{equation}
{\cal L}(\tilde\eta_0,\dots,\tilde\eta_{K-1})={\cal L}[\tilde\eta(\vmin; \vec{\tilde\eta})]
\end{equation}
with $\tilde\eta(\vmin; \vec{\tilde\eta})$ defined in \eqref{eq:discreteEta}.
With this discretization, $\Delta L^c_{\rm min}(v^*,\tilde\eta)$ defined in \eqref{eq:deltaL_c_min} is replaced by a collection of functions $\Delta L_{\rm min}^{c,k}(\tilde\eta^*)$ each having $v^*$ in the $k$-interval $\vmin^k \le v^* < \vmin^{k+1}$, so that $\tilde\eta_k=\eta^*$,
\begin{equation}
  \Delta L_{\rm min}^{c,k}(\tilde\eta^*)
  = -2 \ln \left[ \frac{{\cal L}(\widehat{\widehat{\tilde\eta}}_0,\dots,\widehat{\widehat{\tilde\eta}}_{k-1}, \tilde\eta_{k}=\tilde\eta^*, \widehat{\widehat{\tilde\eta}}_{k+1}, \dots,\widehat{\widehat{\tilde\eta}}_{K-1})}
    {{\cal L}(\widehat{\tilde\eta}_0,\dots,\widehat{\tilde\eta}_{k},\dots,\widehat{\tilde\eta}_{K-1})} \right].
  \label{eq:discDL}
\end{equation}
Here the $\widehat{\widehat{\tilde\eta}}_i$ values maximize the function $\cal L$ subject to the constraint $\tilde\eta_{k}=\tilde\eta^*$, while $\widehat{\tilde\eta}_i$ maximize the function $\cal L$ without the constraint.
Thus $\Delta L_{\rm min}^{c,k}$ is $-2 \ln$ of the profile likelihood ratio (see e.g.~equation (38.53) of \cite{Agashe:2014kda}) with only one parameter $\tilde\eta_k=\tilde\eta^*$.
Notice that the continuous parameter $v^*$ becomes the discrete index $k$, and is no longer an additional parameter.
According to Wilks' theorem, the distribution of $\Delta L_{\rm min}^{c,k}$ approaches a chi-squared distribution with one degree of freedom, in the limit where the data sample is very large \cite{Wilks:1938dza, Agashe:2014kda} (and this is independent of the value of $K$).
In short, this amounts to profiling the likelihood at fixed $v^*$ over the nuisance parameters $\tilde\eta_0, \dots, \tilde\eta_{k-1}, \tilde\eta_{k+1}, \dots, \tilde\eta_K$.
In this language, the fact that the likelihood ratio in \eqref{eq:discDL} has one degree of freedom is proven mathematically in corollary 2 of \cite{Murphy:2000} even for the case $K\to\infty$.

By taking large enough $K$, we can make $\vmin^k$ and $\vmin^{k+1}$ arbitrarily close to $v^*$, and for each $v^*$, $\Delta L_{\rm min}^{c,k}(\tilde\eta^*)$ approaches $\Delta L_{\rm min}^c(v^*, \tilde\eta^*)$.
Therefore, the natural interpretation of the band is the collection of the confidence intervals in $\tilde\eta$ for each $\vmin$ value, which defines a pointwise confidence band, based on a profile likelihood ratio with one degree of freedom.
With this interpretation, we can now compare the confidence band with other limits or measurements in a statistically meaningful way.
If any upper limit at some CL crosses the lower boundary of the band, at some other CL, it means that the two data, providing the limit and the band, are incompatible at their respective CLs.

The Wilks theorem ensures the asymptotic behavior of the distribution of $\Delta L_{\rm min}^c$ as the number of events becomes large, and the $3$ observed number in CDMS-II-Si may not be a large enough number to ensure that $\Delta L$ follows the classical chi-squared distribution.
Assuming that $\Delta L_{\rm min}^c$ is chi-squared distributed, the choices of $\Delta L^*=1.0$ and $\Delta L^*=2.7$ correspond to the confidence intervals of $\tilde\eta$ at the $68\%$ and $90\%$ CL, respectively, for each $\vmin$.
The question on the convergence to the true confidence interval is also present in the analysis of the CDMS-II-Si data with a fixed halo model, if one uses the confidence interval estimator derived from the same likelihood function \cite{Frandsen:2013cna, Gelmini:2014psa}.

In \cite{Fox:2014kua}, $\Delta L^*=9.2$ is used to compute the confidence band at the $90\%$ CL, a value much larger than our choice, corresponding to the $90\%$ CL limit for a chi-squared distribution with five degrees of freedom, resulting from a numerical Monte Carlo simulation.
However, in the simulation described in \cite{Fox:2014kua}, only fake data with three simulated events are selectively generated instead of allowing for any number of simulated events, as would be necessary to
avoid generating a biased data set. 
Yet, allowing the number of simulated events to vary does not seem compatible with the $\Delta L$ definition in Eq.~(2.16) of \cite{Fox:2014kua}.
In this equation, $\sqrt{\Delta L}$ is defined as the radius of a hyper-ellipsoid in a $6$-dimensional parameter space defined by the positions and heights of the three steps in the best-fit $\tilde\eta_{BF}$ for a number of simulated events $N_O=3$. 
This leads to a chi-squared distribution for $\Delta L$ with $2N_O-1=5$ degrees of freedom (because there is one constraint). 
Allowing the number of simulated events $N_O$ to change, the dimension of the $\tilde\eta_{BF}$ parameter space is not fixed to $6$, but would be $2N_O$, leading to a number of degrees of freedom $2N_O-1$ that would change from simulated set to simulated set.

\section{Application to the CDMS-II-Si data}\label{sec:apps}

In this section, we apply the EHI method to the three events observed by CDMS-II-Si in their signal region with recoil energies $8.2$, $9.5$, and $12.3$ keV.
We follow the procedure developed above.
We use $\Delta L^*=1.0$ and $2.7$ for the $68\%$ CL and $90\%$ CL confidence bands, and compare the bands with the $90\%$ CL upper limits from CDMSlite \cite{Agnese:2013lua}, SuperCDMS \cite{Agnese:2014aze}, LUX \cite{Akerib:2013tjd}, XENON100 \cite{Aprile:2012nq} data, as well as the CDMS-II-Si data itself.
The data analysis to obtain the upper limits in this paper is the same as in \cite{Gelmini:2014psa}.
Recent analyses of the CDMS-II-Ge data \cite{Agnese:2014xye,Agnese:2015ywx} use the same data set of \cite{Ahmed:2010wy}, shown in \cite{DelNobile:2014sja} to provide weaker upper limits in the halo-independent analysis than SuperCDMS (and thus not included here).

The data analysis described in this paper is implemented in the CoddsDM software \cite{Georgescu:CoddsDM}, an open-source Python program for comparing the data from  direct detection experiments.

\subsection{Elastic SI scattering}

\begin{figure}[t]
\centering
\includegraphics[width=0.48\textwidth]{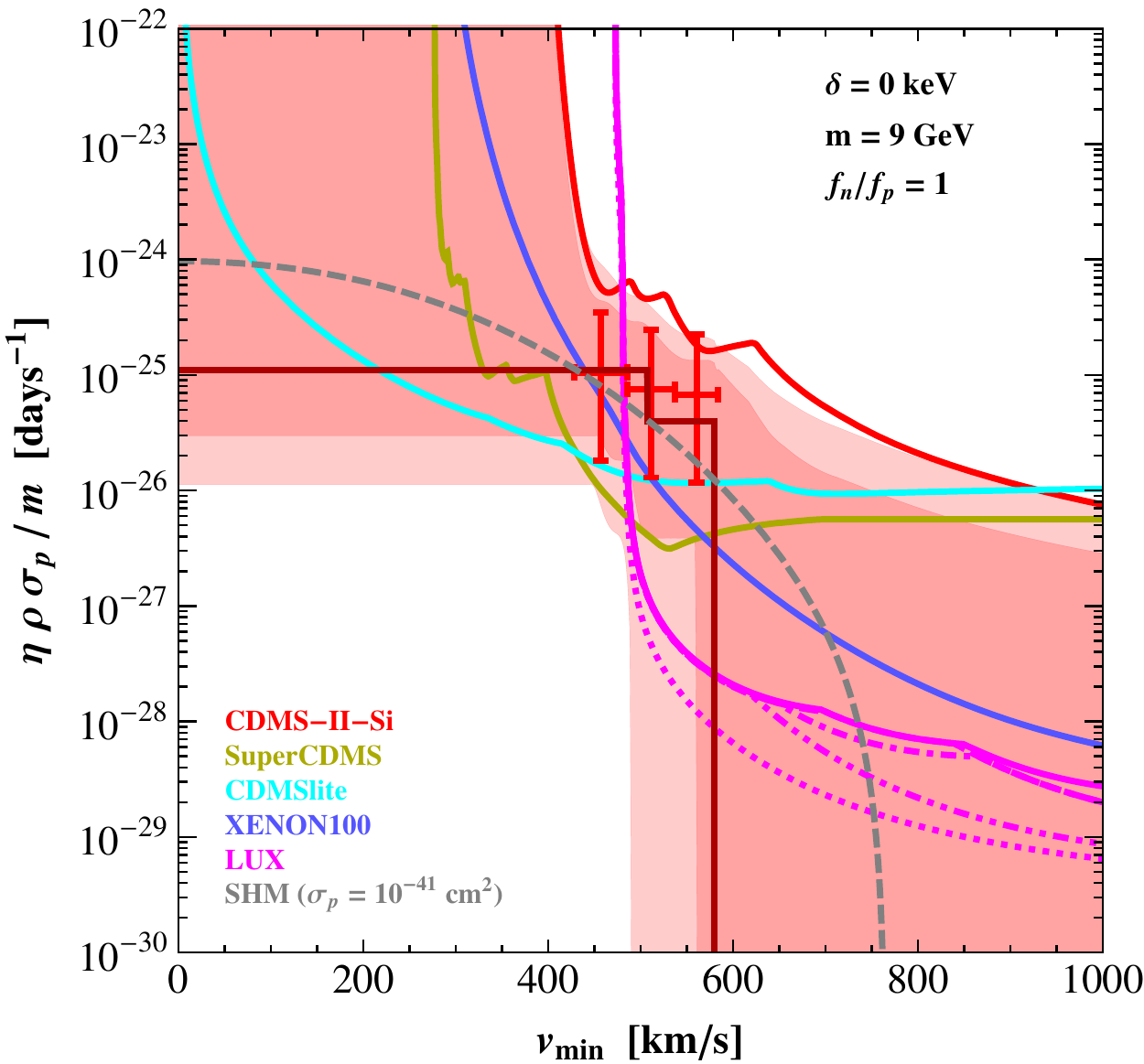}
\includegraphics[width=0.48\textwidth]{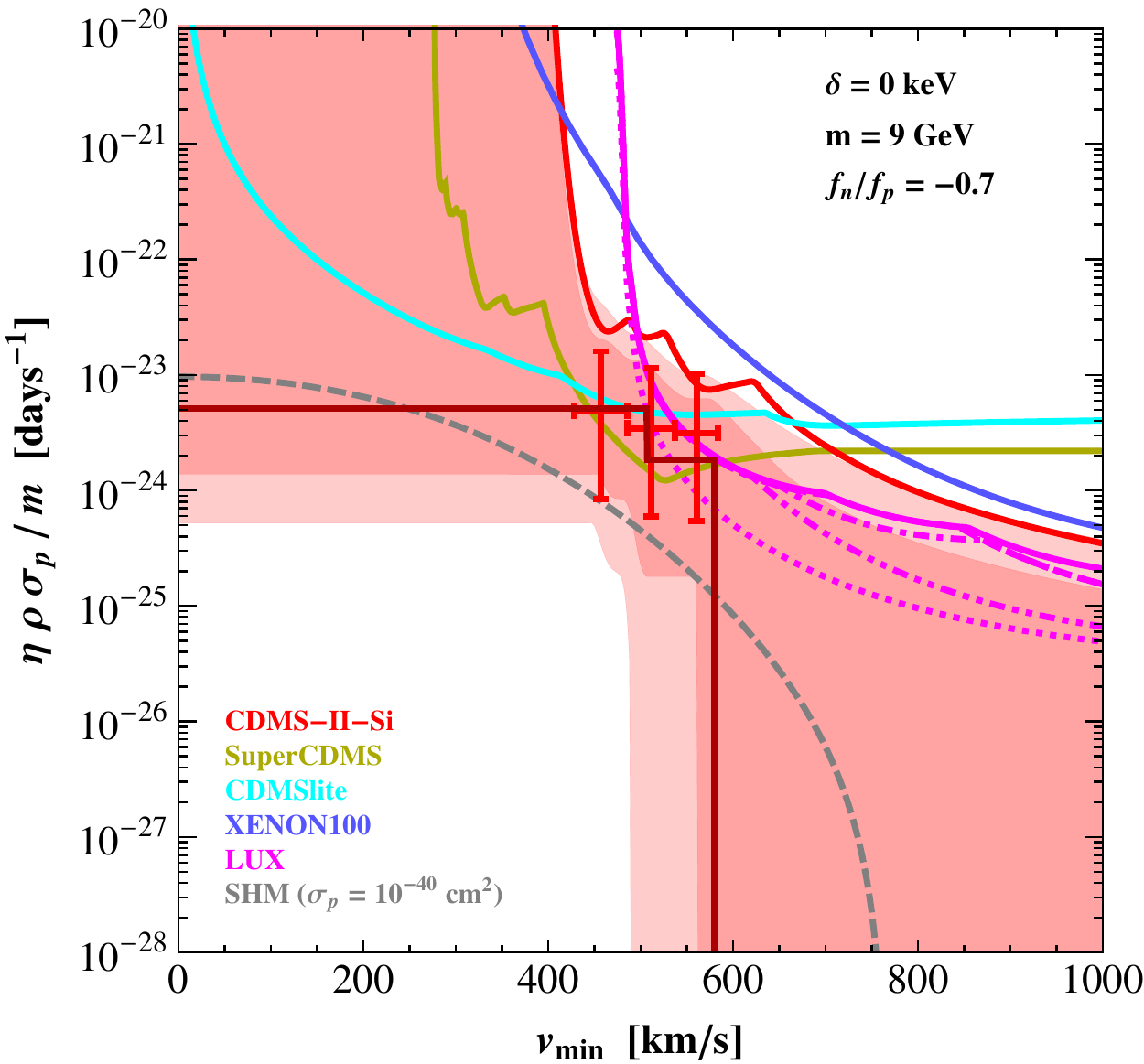}
\caption{90\% CL bounds from CDMSlite, SuperCDMS, LUX, XENON100 and CDMS-II-Si, and the 68\% CL and 90\% CL confidence bands (see the text) from the three WIMP candidate events observed in CDMS-II-Si, for elastic isospin-conserving SI interaction ($f_n/f_p=1$, left panel) and for elastic Xe-phobic isospin-violating SI interaction ($f_n/f_p=-0.7$, right panel), for WIMP mass $m=9$ GeV.}
\label{fig:bandElastic}
\end{figure}

\begin{figure}[t]
\centering
\includegraphics[width=0.48\textwidth]{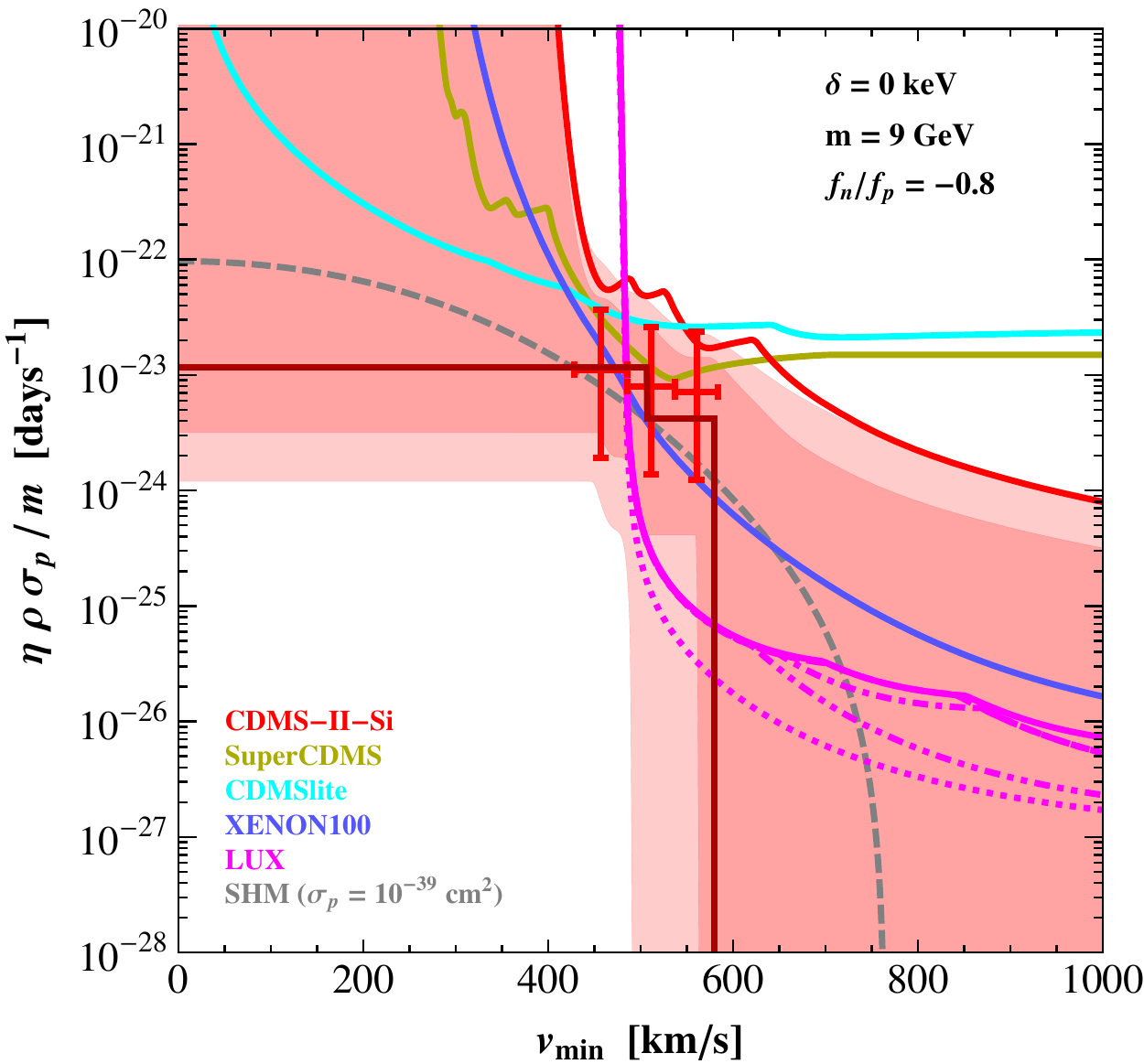}
\caption{Same as Fig.~\ref{fig:bandElastic}, but for elastic Ge-phobic ($f_n/f_p=-0.8$) isospin-violating SI interaction.}
\label{fig:bandElastic_Gephobic}
\end{figure}

\begin{figure}[t]
\centering
\includegraphics[width=0.52\textwidth]{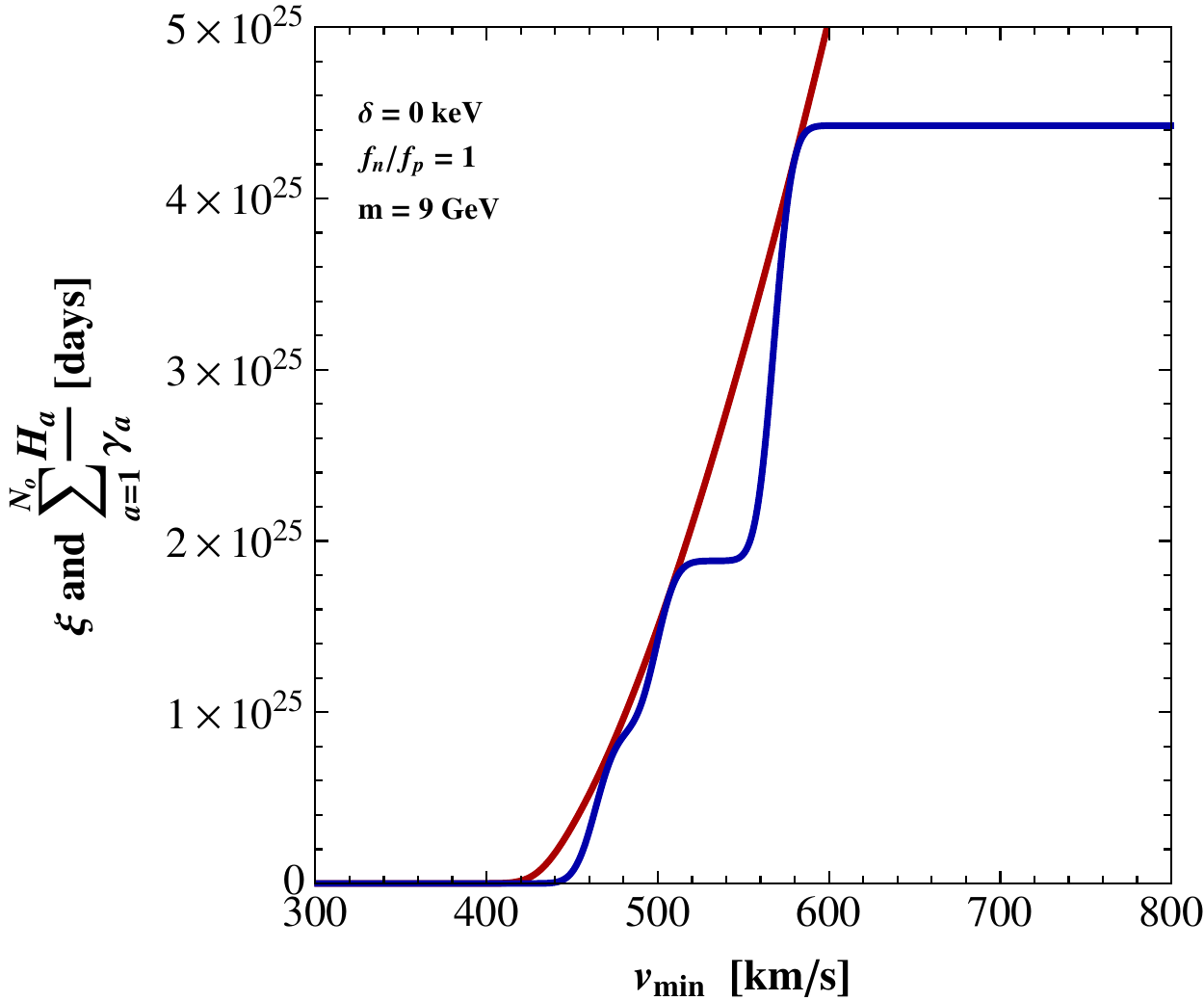}
\includegraphics[width=0.47\textwidth]{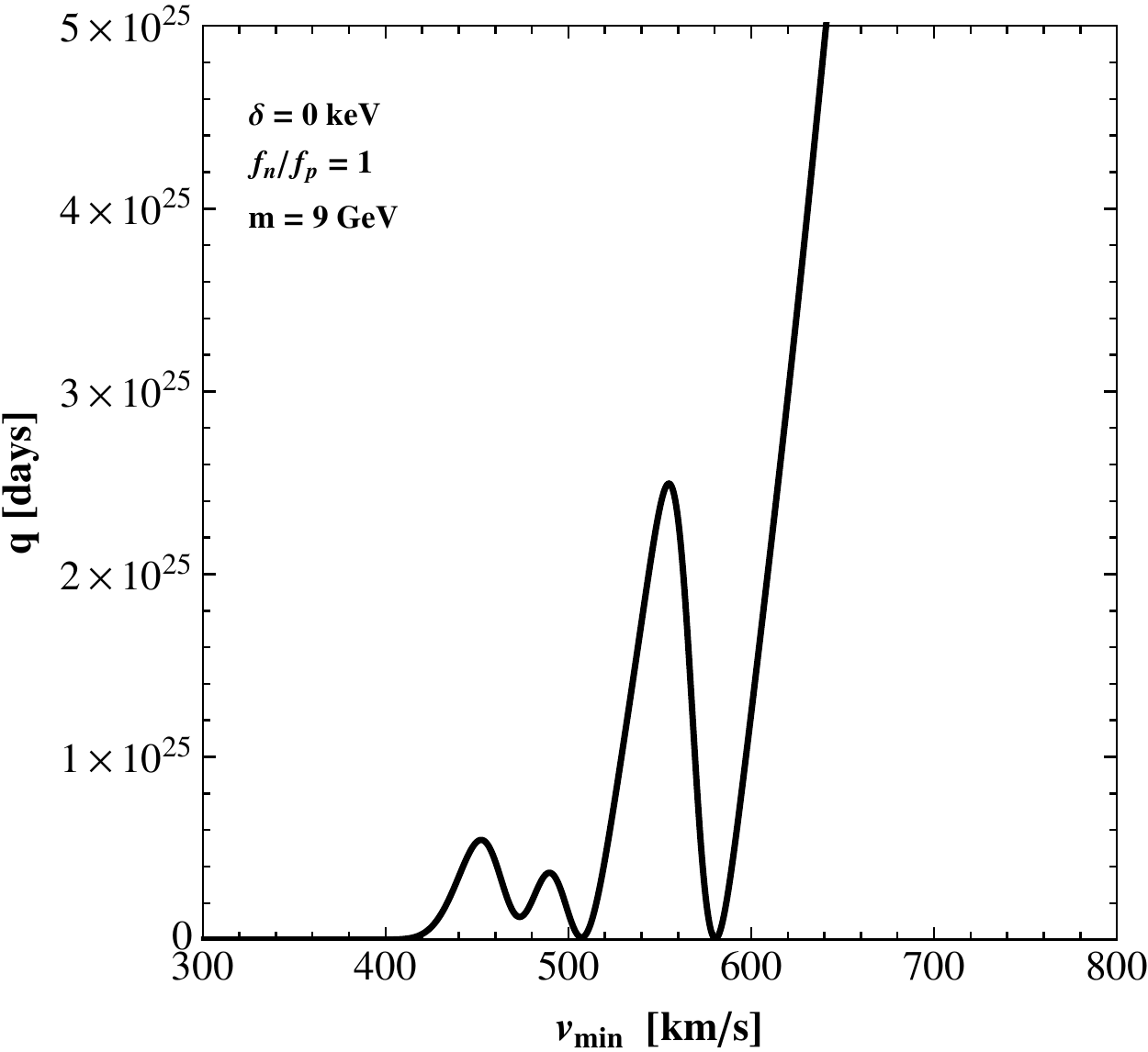}
\caption{$\xi(v_{\rm min})$ (red) and $\sum_{a=1}^{N_O}H_a(v_{\rm min})/\gamma_a$ (blue) (left panel), and $q(v_{\rm min})=2\xi(v_{\rm min})-2\sum_{a=1}^{N_O}H_a(v_{\rm min})/\gamma_a$ (right panel) for SI elastic interaction with $m=9$ GeV (see the text).}
\label{fig:qvdelta0}
\end{figure}

In this subsection we present the result of our analysis for elastic scattering with isospin-conserving $f_n/f_p=1$ and with isospin-violating $f_n/f_p=-0.7$ (Xe-phobic) and $f_n/f_p=-0.8$ (Ge-phobic) SI interactions \cite{Kurylov:2003ra,Chang:2010yk,Feng:2011vu}.
They are shown in the left and right panels of Fig.~\ref{fig:bandElastic} and in Fig.~\ref{fig:bandElastic_Gephobic}, respectively, for a WIMP of mass $m=9$ GeV.
This value of the mass is within the $68\%$ CL CDMS-II-Si regions obtained assuming the Standard Halo Model (SHM) in \cite{DelNobile:2014sja} and \cite{Gelmini:2014psa}.
Figs.~\ref{fig:bandElastic} and \ref{fig:bandElastic_Gephobic} show the best-fit $\tilde\eta_\text{BF}$ (dark red line) and the $68\%$ and $90\%$ CL confidence bands derived from the CDMS-II-Si data shaded in darker and lighter red, respectively.
Despite starting with three observed events, thus three steps in $\tilde\eta$, the $\tilde\eta_\text{BF}$ has only two steps, located at the zeros of the $q(\vmin)$ function shown in Fig.~\ref{fig:qvdelta0}.

Fig.~\ref{fig:qvdelta0} shows the $\xi(\vmin)$ (red lines) and $\sum_a H_a(\vmin)/\gamma_a$ (blue lines) functions in the left panel, and the $q(\vmin)$ function given in \eqref{eq:qv} (right panel) for the best-fit $\tilde\eta_{\rm BF}$ of the CDMS-II-Si data for spin-independent elastic scattering with $f_n/f_p=1$.
The zeros of $q(\vmin)$, located at the points where the functions in the left panel of Fig.~\ref{fig:qvdelta0} touch, are at $507$ and $580$ km/s.
These coincide with the locations of the steps of the best-fit $\tilde\eta_{\rm BF}$ plotted in the left panel of Fig.~\ref{fig:bandElastic}.
The location of the steps is practically the same for other $f_n/f_p$ values.
The shapes of the $\xi(\vmin)$, $\sum_a H_a(\vmin)/\gamma_a$, and $q(\vmin)$ functions are almost unchanged for a different choice of $f_n/f_p$ values, up to a rigid rescaling along the vertical axis.
The only changes expected in the positions of the zeros of $q(\vmin)$ for different $f_n/f_p$ values are due to the very small change in the relative strength of the WIMP interaction with different isotopes.
Figs.~\ref{fig:bandElastic} and \ref{fig:bandElastic_Gephobic} show the $90\%$ CL CDMSlite (cyan), SuperCDMS (dark yellow), LUX (magenta), XENON100(blue) and CDMS-II-Si (red) upper limits, and the red crosses derived from the halo-independent analysis using binned data \cite{DelNobile:2014sja}.
The crosses represent $68\%$ CL intervals of averaged $\tilde\eta$ and the corresponding $\vmin$ ranges for the CDMS-II-Si data with three equally spaced bins spanning the recoil energy range from $7$ to $13$ keV.
Notice that the $68\%$ CL crosses are similar in vertical extent to the $68\%$ CL confidence band.
Notice also that the $90\%$ CL CDMS-II-Si limit follows closely the upper limit of the $90\%$ CL confidence band.

As one can see in the left panel of Fig.~\ref{fig:bandElastic} for $f_n/f_p=1$, the $68\%$ CL confidence band is excluded in the $\vmin$ range from $370$ to $560$ km/s, by the combination of the $90\%$ CL CDMSlite, SuperCDMS, and LUX upper limits.
The lower boundary of the $90\%$ CL confidence band is also cut at $450$ km/s by the SuperCDMS $90\%$ CL limit.
Since there is no single continuous curve within the $90\%$ CL confidence band which does not cross any $90\%$ CL upper limit, we conclude that the potential signal and limits are incompatible for any halo model.

On the other hand, in the right panel of Fig.~\ref{fig:bandElastic} a significant portion of the $68\%$ CL confidence band remains below all the $90\%$ CL upper limits.
This shows that for SI interactions with $f_n/f_p=-0.7$ the CDMS-II-Si signal is consistent with the null results of all other experiments.

The choice of $f_n/f_p=-0.8$ (Fig.~\ref{fig:bandElastic_Gephobic}) disfavors maximally the Ge limits (while $f_n/f_p=-0.7$ disfavors maximally Xe couplings instead).
Thus, as expected, in Fig.~\ref{fig:bandElastic_Gephobic} the SuperCDMS limit is weakened with respect to Fig.~\ref{fig:bandElastic}, but the LUX upper limits exclude almost completely both confidence bands.

The dashed gray curves in Figs.~\ref{fig:bandElastic} and \ref{fig:bandElastic_Gephobic} are the $\tilde\eta$ functions assuming the SHM for WIMP-proton cross sections $\sigma_p=10^{-41}$ cm$^2$ and $\sigma_p=10^{-40}$ cm$^2$ in the left and right panels of Fig.~\ref{fig:bandElastic}, and $\sigma_p=10^{-39}$ cm$^2$ in Fig.~\ref{fig:bandElastic_Gephobic}.
For $m=9$ GeV, these $\sigma_p$ values are within the $68\%$ and $90\%$ CL CDMS-II-Si regions obtained assuming the SHM (in Fig.~1 of \cite{DelNobile:2014sja} and in Fig.~4 of \cite{Gelmini:2014psa}, respectively).
In the analyses of \cite{DelNobile:2014sja} and \cite{Gelmini:2014psa} assuming the SHM, the $m$ and $\sigma_p$ choices for $f_n/f_p=1$ and $f_n/f_p=-0.8$ interactions are shown to be rejected, while the choice for $f_n/f_p=-0.7$ interactions are allowed by all $90\%$ upper limits.
The same conclusions are evident in Figs.~\ref{fig:bandElastic} and \ref{fig:bandElastic_Gephobic}, where the dashed gray lines are above the upper limits for $f_n/f_p=1$ and $f_n/f_p=-0.8$ and below them for $f_n/f_p=-0.7$.

\subsection{Inelastic SI scattering}

\begin{figure}[t]
\includegraphics[width=0.48\textwidth]{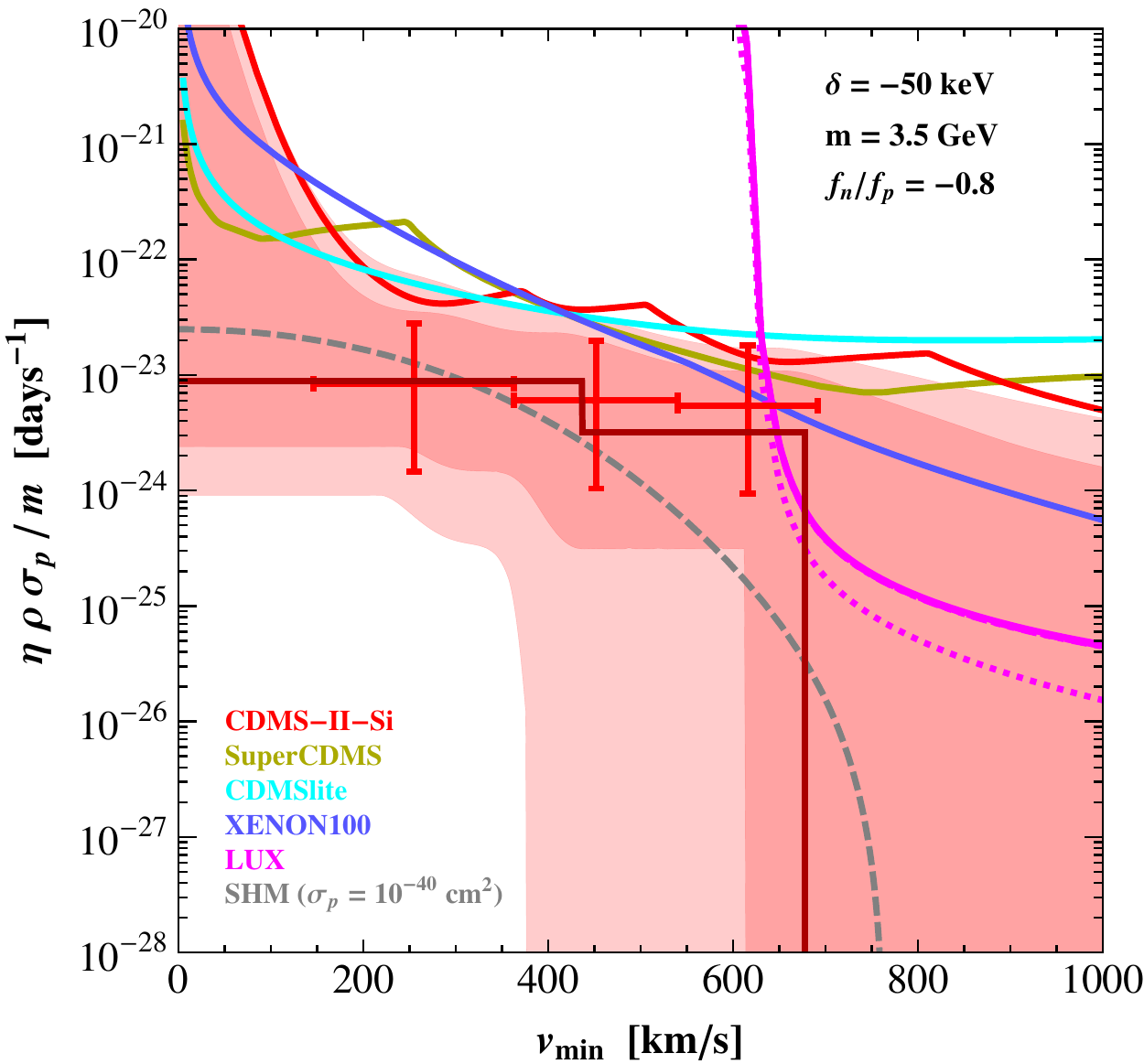}
\includegraphics[width=0.48\textwidth]{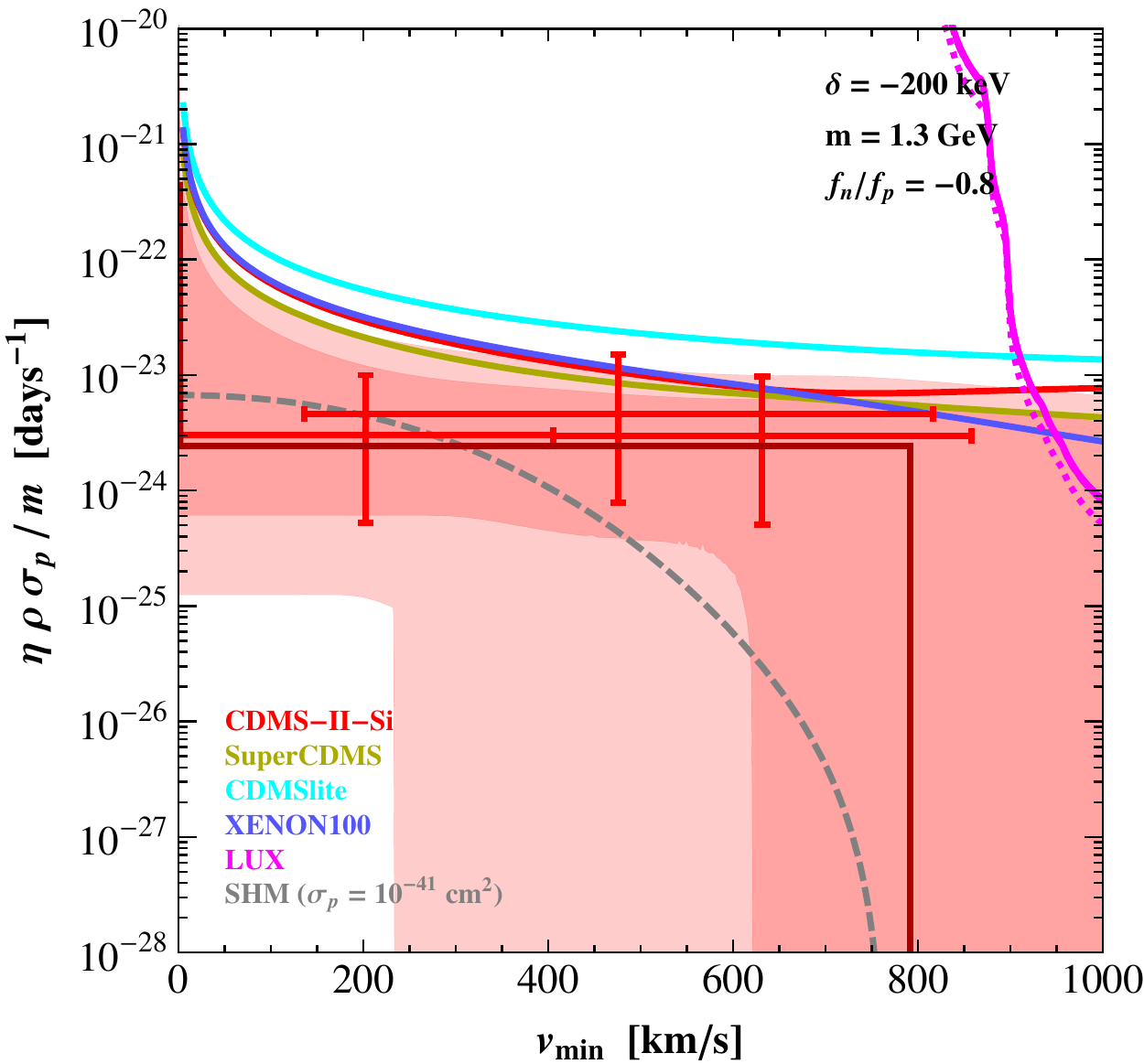}
\caption{Same as Fig.~\ref{fig:bandElastic}, but for Ge-phobic isospin-violating SI interaction ($f_n/f_p=-0.8$) with $m=3.5$ GeV and $\delta=-50$ keV (left panel), and $m=1.3$ GeV and $\delta=-200$ keV (right panel).}
\label{fig:bandGePhobic}
\end{figure}

In this subsection we present the results of the analysis for the exothermic Ge-phobic WIMP proposed in \cite{Gelmini:2014psa,Scopel:2014kba} as an interpretation of the CDMS-II-Si data, shown in Fig.~\ref{fig:bandGePhobic}.This choice of $f_n/f_p=-0.8$ suppresses maximally the coupling to Ge.
The limits due to Xe are weakened by the exothermic nature of the scattering, which disfavors heavier targets (such as Xe) with respect to lighter ones (such as Si) \cite{Gelmini:2014psa}, leaving in principle Ge limits as the most important.

Fig.~\ref{fig:bandGePhobic} shows our results for a WIMP mass $m=3.5$ GeV and mass split $\delta=-50$ keV (left panel), and $m=1.3$ GeV and $\delta=-200$ keV (right panel).
These masses are shown in \cite{Gelmini:2014psa} to be within the CDMS-II-Si $90\%$ and $68\%$ CL regions when assuming the SHM, for $\sigma_p=10^{-40}$ cm$^2$ and $10^{-41}$ cm$^2$, respectively (see Figs.~5 and 6 of \cite{Gelmini:2014psa}).
This is corroborated by the present halo-independent analysis, since the corresponding $\tilde\eta$ functions assuming the SHM shown in Fig.~\ref{fig:bandGePhobic} (dashed gray lines) escape all upper limits from experiments with null results.
\begin{figure}[t]
\centering
\includegraphics[width=0.52\textwidth]{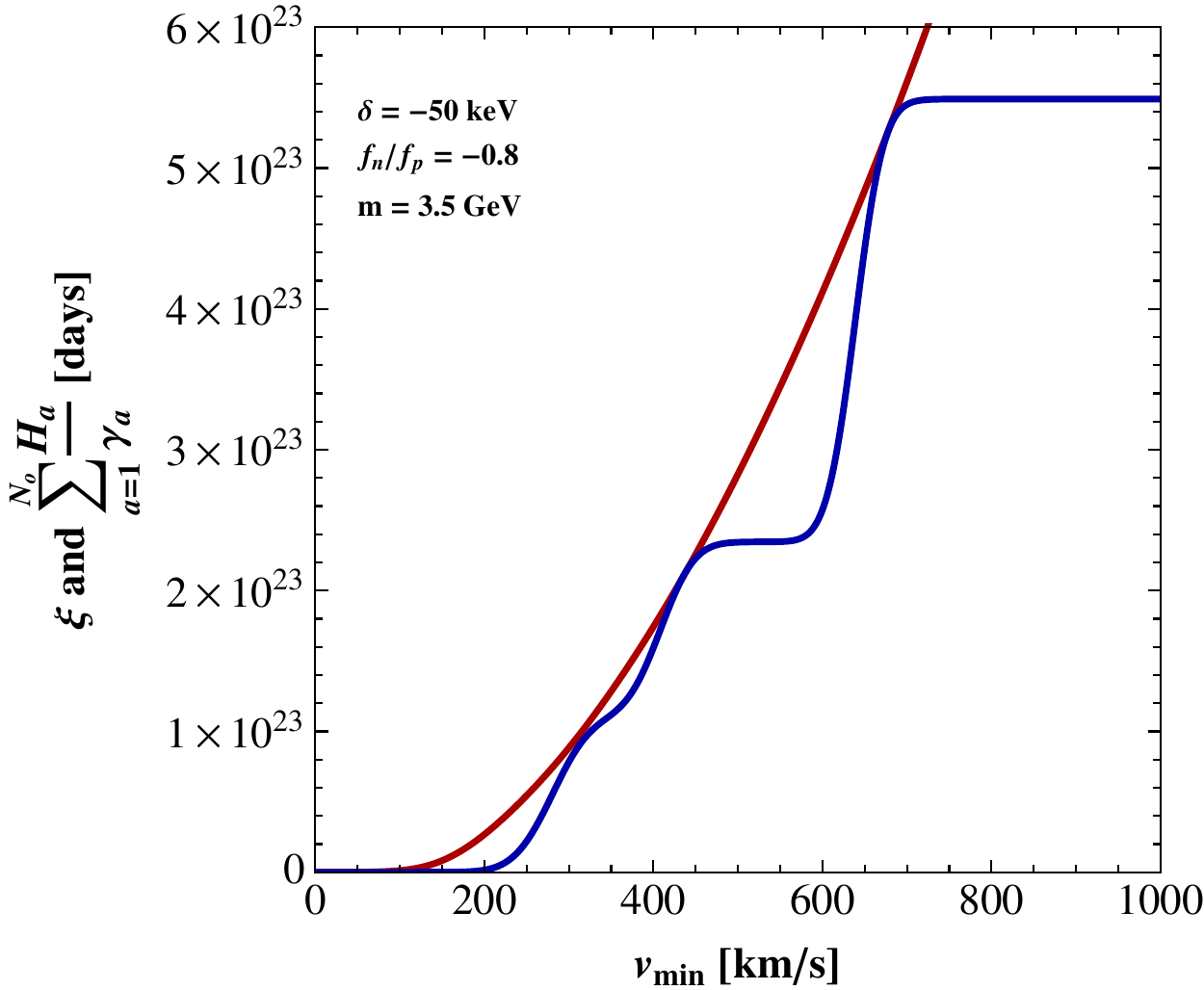}
\includegraphics[width=0.47\textwidth]{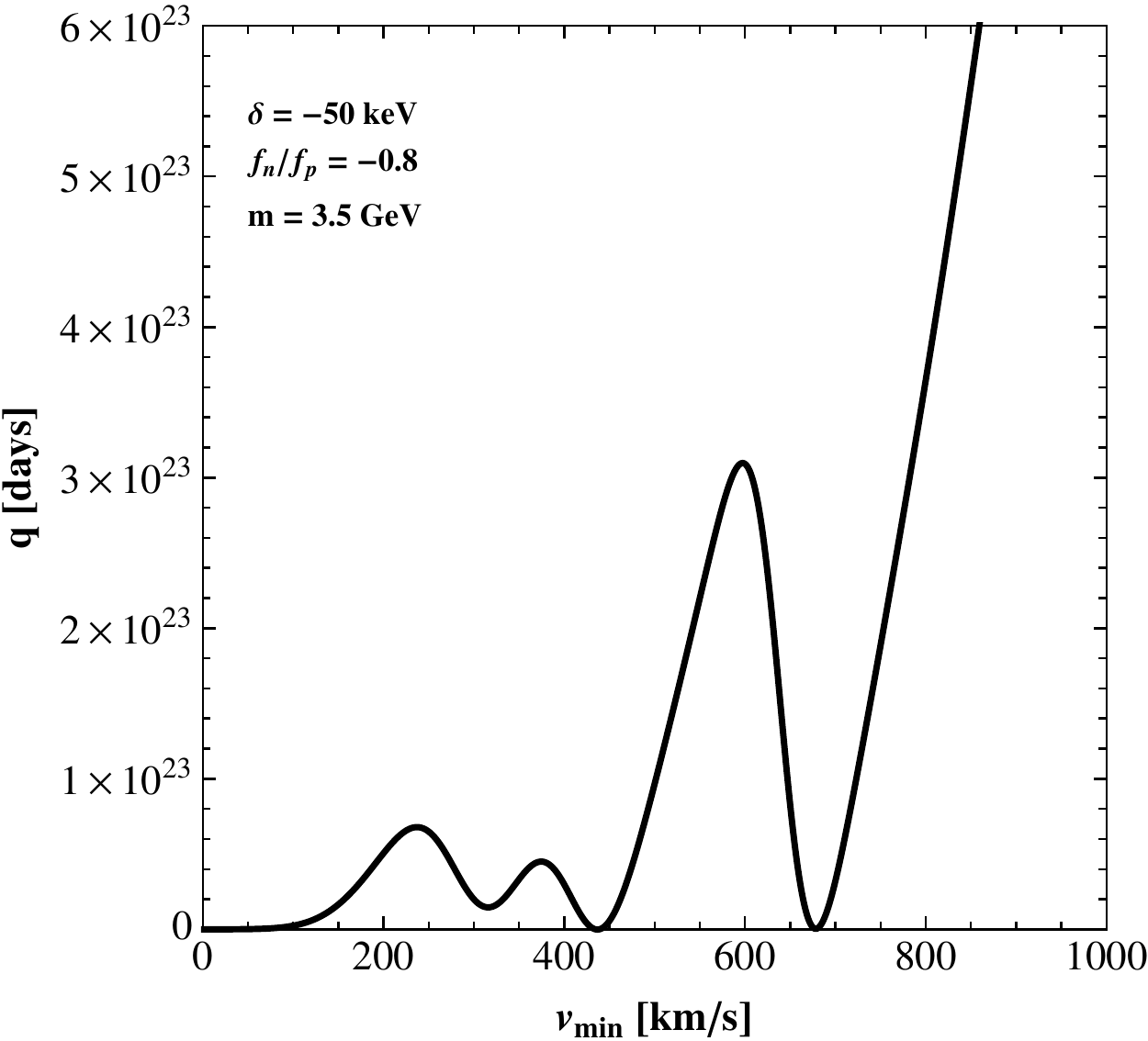}
\caption{Same as Fig.~\ref{fig:qvdelta0}, but for SI exothermic inelastic interaction with $m=3.5$ GeV and $\delta=-50$ keV (see the text).}
\label{fig:qvdeltaneg50}
\end{figure} 
\begin{figure}[t]
\centering
\includegraphics[width=0.52\textwidth]{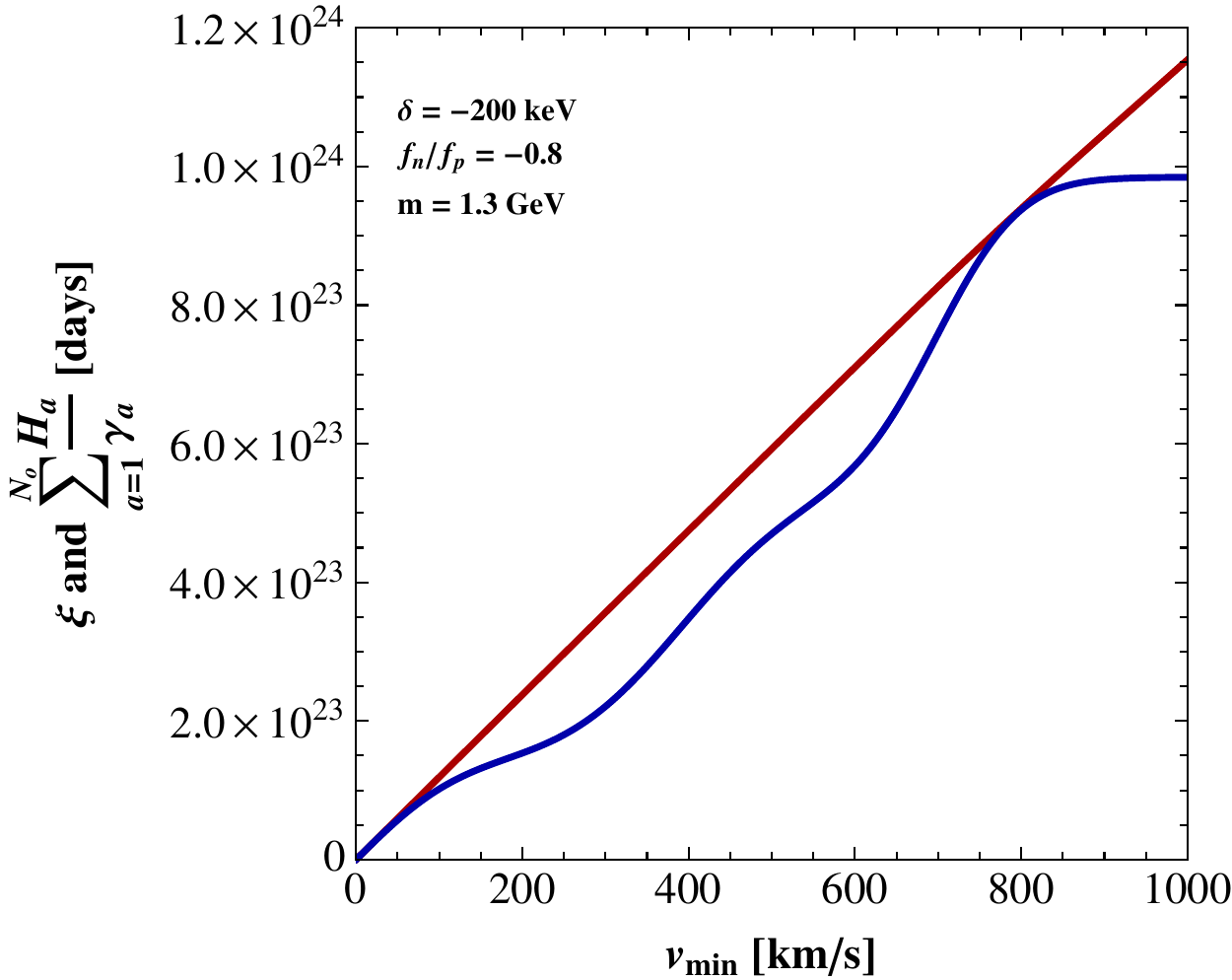}
\includegraphics[width=0.47\textwidth]{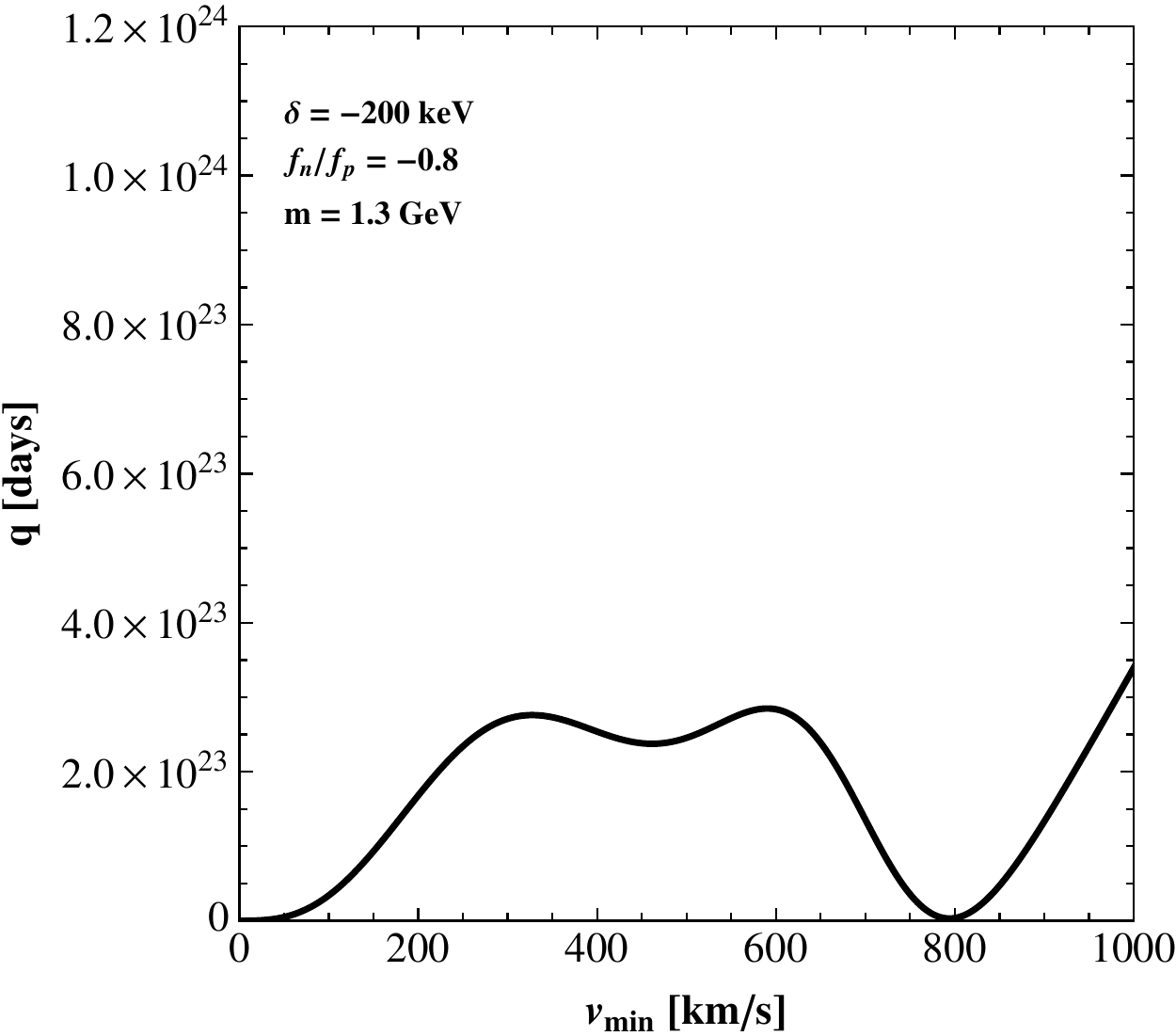}
\caption{Same as Fig.~\ref{fig:qvdelta0}, but for SI exothermic inelastic interaction with $m=1.3$ GeV and $\delta=-200$ keV (see the text).}
\label{fig:qvdeltaneg200}
\end{figure} 

The best-fit $\tilde\eta_\text{BF}$ functions for both Ge-phobic cases are shown in dark red in Fig.~\ref{fig:bandGePhobic}.
They have two and one steps respectively in the left and right panels of Fig.~\ref{fig:bandGePhobic}, corresponding to the zeros of the $q(\vmin)$ functions in the right panels of Figs.~\ref{fig:qvdeltaneg50} and \ref{fig:qvdeltaneg200} (located at $\vmin$ values of $437$ and $678$ km/s in Fig.~\ref{fig:qvdeltaneg50} and $792$ km/s in Fig.~\ref{fig:qvdeltaneg200}).

Figs.~\ref{fig:qvdeltaneg50} and \ref{fig:qvdeltaneg200} show the functions $\xi$ (red) and $\sum_{a=1}^{N_O}H_a(v_{\rm min})/\gamma_a$ (blue) in the left panels, and twice their difference, $q(\vmin)$, in the right panels, for the two Ge-phobic cases in Fig.~\ref{fig:bandGePhobic}.

In the previous analysis of Ge-phobic exothermic WIMP based on the SHM \cite{Gelmini:2014psa}, the $m$ and $\sigma_p$ parameters chosen in the current analysis are found to be compatible with the null results of all other experiments.
Consistently with this result, we find a large portion of the $68\%$ CL confidence band is below all the $90\%$ CL upper limits imposed by all null results.
Thus WIMP-nucleus scattering through Ge-phobic interaction can potentially explain the CDMS-II-Si data as a WIMP signal without any conflict with the null results of all other searches.

\section{Conclusions}\label{sec:con}

We have expanded and corrected a recently proposed extended maximum likelihood halo-independent (EHI) method to analyze unbinned direct dark matter detection data.
Instead of the recoil energy $E_R$ as independent variable, we use $\vmin$, the minimum speed a dark matter particle must have to impart a given recoil energy to a nucleus.
An earlier version of the method, using $E_R$ as variable, was introduced in \cite{Fox:2014kua}.
The use of $\vmin$ as variable allows to incorporate in the analysis any type of target composition and of WIMP-nucleus interaction, including elastic and inelastic collisions.
This is not possible using $E_R$.
The advantages of using $\vmin$ instead of $E_R$ in a halo-independent analysis was first pointed out in \cite{Gondolo:2012rs} and extensively used later on \cite{DelNobile:2013cta,DelNobile:2013cva,Gelmini:2014psa,DelNobile:2014eta,DelNobile:2014sja}.

The EHI method uses unbinned direct dark matter detection data.
The predicted differential rate as a function of the observed energy $E'$ in all direct detection experiments can be written in terms of a common function $\tilde\eta(\vmin)$ (see \eqref{eq:rateInCalR}).
The aim of the method is to find the $\tilde\eta$ function that provides the best fit for the unbinned data.
We have proven rigorously that the best-fit $\tilde\eta$ function, $\tilde\eta_\text{BF}(\vmin)$, is a piecewise constant function with a number of discontinuities smaller than or equal to the number of observed events $N_O$.
We have also rigorously defined a two-sided pointwise confidence band with a clear statistical meaning, as a collection of confidence intervals in $\tilde\eta$ for every $\vmin$ value.
We can assign a confidence level to the band and thus compare with upper limits given at particular confidence levels.
This allows to quantitatively assess the compatibility of the unbinned data with upper limits due to null results.

Using this method, we analyzed the compatibility of the three candidate events found by CDMS-II-Si with the best available upper bounds, for spin-independent (SI) WIMP-nucleus interactions with different neutron to proton coupling ratio $f_n/f_p$ values and either elastic or exothermic inelastic scattering.
We found the best-fit $\tilde\eta_\text{BF}$ function and $68\%$ and $90\%$ CL confidence bands.
We chose values of the WIMP mass within the CDMS-II-Si regions in the $m$--$\sigma_p$ plane that we had found in previous studies \cite{Gelmini:2014psa,DelNobile:2014sja} assuming the Standard Halo Model (SHM).
Our results for $f_n/f_p=1$, WIMP mass $m=9$ GeV and elastic scattering are shown in the left panel of Fig.~\ref{fig:bandElastic}.
The $90\%$ CL CDMSlite, SuperCDMS and LUX limits derived as in \cite{Gelmini:2014psa} exclude the entire $90\%$ CL band for this candidate.
This case was also studied in \cite{Fox:2014kua}, where the best-fit $\tilde\eta_\text{BF}$ is very similar to ours, but the $90\%$ CL band is much larger.
In \cite{Fox:2014kua}, only $90\%$ CL limits derived from LUX and XENON10 data are presented.
The LUX limit in \cite{Fox:2014kua} is similar to ours, but it does not exclude their much larger confidence band.

The right panel of Fig.~\ref{fig:bandElastic} shows our results for $f_n/f_p=-0.7$ (Xe-phobic) and $m=9$ GeV.
We found that in this case a significant portion of the $68\%$ and $90\%$ CL confidence bands remains below all the $90\%$ CL upper limits.
Thus, a WIMP candidate with these characteristics provides an explanation for the three CDMS-II-Si events compatible with all present null results of other direct searches.
This case was also studied in \cite{Fox:2014kua}, where their best-fit $\tilde\eta_\text{BF}$ function has the same number and position of steps as ours, but is an order of magnitude larger.
We think this difference {\color{blue}might} be due, {\color{blue} at least in part}, to the inclusion of the isotopic composition of Si in our computation, which can not be done with the method used in \cite{Fox:2014kua}.
The LUX limit presented in \cite{Fox:2014kua} for this case is similar to ours, but their $90\%$ CL band is again much larger.

The Ge-phobic $f_n/f_p=-0.8$ case, again for $m=9$ GeV, is presented in Fig.~\ref{fig:bandElastic_Gephobic}.
The $90\%$ CL confidence band is almost completely excluded by the $90\%$ CL LUX limit.

Our results for the Ge-phobic coupling and exothermic inelastic scattering are presented in Fig.~\ref{fig:bandGePhobic}, for two different values of the WIMP mass $m$ and mass split $\delta$: $m=3.5$ GeV, $\delta=-50$ keV and $m=1.3$ GeV, $\delta=-200$ keV.
In these cases the $68\%$ and $90\%$ CL confidence bands are almost entirely below all the $90\%$ CL limits.
Thus, again we found compatibility between a dark matter interpretation of the CDMS-II-Si data and all null results.

In all cases studied we included the crosses derived from the CDMS-II-Si data obtained with our previous halo-independent analysis using binned data \cite{Gelmini:2014psa,DelNobile:2014sja}.
The crosses represent $68\%$ CL intervals of averaged $\tilde\eta$ and $\vmin$ ranges corresponding to three equally-spaced bins spanning the recoil energy range from $7$ to $13$ keV.
We found that the crosses are similar in vertical extent to the $68\%$ CL confidence bands in all cases.
This shows agreement between both types of halo-independent analyses, but the present method is much more powerful.

We found remarkable that the $90\%$ CL limit derived from the CDMS-II-Si data itself using the Maximum Gap method, as described in \cite{Gelmini:2014psa,DelNobile:2014sja} (and references therein) is almost identical to the $90\%$ CL upper boundary of the $90\%$ CL confidence band in all cases studied.
Again, this indicates agreement between the two different analyses.

SI elastic scattering was also studied in \cite{Feldstein:2014ufa} and \cite{Bozorgnia:2014gsa}, where
two different statistics were used to quantify the compatibility among different direct search data sets.
In \cite{Feldstein:2014ufa}, for isospin-conserving SI interactions and WIMP mass $7$ GeV, which is slightly smaller than our choice of $9$ GeV, the parameter goodness-of-fit value derived from the global likelihood of the CDMS-II-Si, SuperCDMS and LUX data has a p-value of only $0.44\%$.
This poor compatibility level is consistent with our results.
For isospin-violating interactions, \cite{Feldstein:2014ufa} used slightly different parameter sets, $f_n/f_p=-0.71$, $m=6.2$ GeV, and $f_n/f_p=-0.79$, $m=6.3$ GeV, with corresponding p-values of $18.7\%$ and $18.5\%$.
Thus the compatibility is significantly improved, which is also consistent with our results. 
In \cite{Bozorgnia:2014gsa}, a test statistic ``$p_{\rm joint}$'' is proposed and calculated, said to be the upper bound on the joint probability of obtaining the outcomes of two potentially conflicting experiments.
Only if the value of $p_{\rm joint}$ is small there is a clear interpretation of incompatibility, but a large $p_{\rm joint}$ value does not imply compatibility.
For $m=9$ GeV, \cite{Bozorgnia:2014gsa} finds incompatibility between CDMS-II-Si and SuperCDMS for $f_n/f_p=1$, but not for $f_n/f_p=-0.7$ or $-0.8$.
In this respect, we agree.

The use of a test statistic such as defined in \cite{Feldstein:2014ufa} or \cite{Bozorgnia:2014gsa} is complementary to our method of using a confidence band and upper limits in $\vmin-\tilde\eta$ space to assess the compatibility among different data sets.

\section*{Acknowledgments}

The authors thank Eugenio Del Nobile for many helpful discussions.
G.G., A.G.~and J.-H.H.~were supported in part by the Department of Energy under Award Number DE-SC0009937.
J.-H.H.~was also partially supported by Spanish Consolider-Ingenio MultiDark (CSD2009-00064).
P.G.~was supported in part by NSF grant PHY-1415974.

\input{bibliography}

\end{document}

%% file: bibliography.tex
\newpage{\pagestyle{empty}\cleardoublepage}